\documentclass[reprint,aps,pra,amsmath,amssymb,superscriptaddress,nofootinbib,nolongbibliography]{revtex4-2}

\usepackage{graphicx}
\usepackage{bm}
\usepackage{tikz}
\usetikzlibrary{quantikz2,decorations.pathmorphing,shapes.geometric}
\usepackage[colorlinks=true,linkcolor=blue,citecolor=blue,urlcolor=blue]{hyperref}

\AtBeginDocument{%
  \renewcommand{\ket}[1]{|#1\rangle}%
  \renewcommand{\bra}[1]{\langle #1|}%
  \renewcommand{\braket}[2]{\langle #1|#2\rangle}}
\DeclareMathOperator{\tr}{tr}
\newcommand{\Hp}{\mathcal{H}_{\mathrm{p}}}
\newcommand{\Ho}{\mathcal{H}_{\mathrm{o}}}
\newcommand{\He}{\mathcal{H}_{\mathrm{E}}}
\newcommand{\rhop}{\rho_{\mathrm{p}}}
\newcommand{\rhoo}{\rho_{\mathrm{o}}}
\newcommand{\Ip}{I_{\mathrm{p}}}
\newcommand{\Io}{I_{\mathrm{o}}}
\newcommand{\W}{\mathcal{W}}
\newcommand{\st}[1]{\tilde{\sigma}_{#1}}
\newcommand{\norm}[1]{\lVert #1\rVert}
\newcommand{\Pc}{P_{\mathrm{coinc}}}

\begin{document}

\title{Witnessing internal polarization entanglement of single photons\\ with two-photon interference}

\author{Guilherme P. Tempor\~ao}
\email{temporao@puc-rio.br}
\affiliation{NiTeQ, Pontifical Catholic University of Rio de Janeiro, 22451-900 Rio de Janeiro, Brazil}

\date{\today}

\begin{abstract}
A partially polarized photon can result from two distinct mechanisms: the coupling of its polarization to another internal degree of freedom, such as time of arrival or frequency, or the entanglement of its polarization with an external environment. In a previous work [G.~P. Tempor\~ao \emph{et al.}, Phys.\ Rev.\ A \textbf{109}, 022402 (2024)], it was shown that Hong-Ou-Mandel (HOM) interference between two copies of a photon, one of which is subjected to each of the Pauli operators, can distinguish these mechanisms, provided only one of them is present. In this work, we show that the same four measurements can be employed to certify ``internal'' entanglement, even when the photons are also entangled with the environment or are in an incoherent mixture of pure states. By averaging the coincidence probabilities, one of the photons is subject to a completely depolarizing channel, which gives us the purity of the remaining degrees of freedom without resolving them; comparing it with the purity of the whole photon implements a known separability criterion as a linear combination of coincidence probabilities. We specify the conditions for reliable internal entanglement certification and analyze the main experimental imperfections. We apply the method to detection of polarization--frequency entanglement induced by the fine-structure splitting in quantum dots, without temporal or spectral resolution and without a source model. We also show that the method generalizes to qudits.
\end{abstract}

\maketitle

%=====================================================================
\section{Introduction}
%=====================================================================

Partial polarization of a single photon may arise from two different physical mechanisms. Its polarization may be coupled to another internal degree of freedom (DOF), such as its time of arrival, frequency, or transverse mode, or it may be entangled with some external quantum system, i.e., the environment~\cite{Jeffrey2004,Peters2005}. Both situations are common in photonic experiments: the fine-structure splitting of quantum-dot excitons couples polarization to frequency~\cite{Stevenson2006,Akopian2006}, birefringence and polarization-mode dispersion (PMD) in optical fibers couple it to the time of arrival~\cite{GordonKogelnik2000}, and hyperentangled states are engineered such that several DOFs of the same photon are correlated~\cite{Kwiat1997,Barreiro2005}. In the context of single-photon sources (SPSs) aimed at quantum communication, photonic quantum computing, and quantum metrology~\cite{GisinThew2007,OBrien2007,Couteau2023}, being able to distinguish these mechanisms is important, as a coherent coupling between DOFs of the same photon can, in principle, be undone by a suitable unitary transformation, whereas entanglement with the environment cannot. Even though polarization quantum state tomography (QST)~\cite{James2001,Altepeter2005} can be employed to characterize partial polarization, it is blind to this distinction: the reduced density operator does not contain information on the DOFs that have been traced over unless these DOFs are independently resolved.

Two-photon interference, namely the Hong-Ou-Mandel (HOM) effect~\cite{HOM1987}, however, may allow such distinction. The HOM effect, known to be equivalent to the swap test (ST)~\cite{Buhrman2001,GarciaEscartin2013}, allows direct measurement of nonlinear functionals of a quantum state, such as its purity~\cite{Ekert2002,Adamson2007}. Such two-copy measurements have been employed to quantify the entanglement of photon pairs~\cite{Bovino2005,Walborn2006} and of many-body systems~\cite{Daley2012,Islam2015}, and they can even be exploited to provide exponential advantage in some learning tasks~\cite{Huang2022}. In a previous work~\cite{Temporao2024}, we have shown that, if one of the two interfering photons is subjected to each of the four Pauli operators (including the identity), the measured coincidence probabilities yield the absolute values of the Stokes parameters and the DOP, without resorting to any polarization-selective component. Moreover, they reveal whether the photon is in a pure state; if so, any depolarization must be due to ``internal'' entanglement; otherwise, the depolarization was attributed to ``external'' entanglement or to incoherent mixtures of pure states.

However, this reasoning only holds when one of the two mechanisms is absent. In a realistic source, both of them may be present at the same time; for instance, a quantum-dot photon whose polarization is coupled to its frequency through the fine-structure splitting will also be partially mixed due to phonon-induced dephasing. In such a case, the photon is in a mixed state, and the criterion of Ref.~\cite{Temporao2024} cannot tell whether internal entanglement is present or not.

In this work, we show that the four measurements of Ref.~\cite{Temporao2024} already contain enough information to ascertain the existence of internal entanglement. By averaging the HOM coincidence probabilities over the four Pauli operators, the photon undergoing the operations is depolarized, such that its polarization is ``erased'' from the interference, resulting in a ST probing only the remaining DOFs. A linear combination of the four coincidence probabilities gives rise to a quantity that never takes positive values unless the polarization is entangled with another internal DOF, irrespective of the amount of external entanglement or classical mixing.

We show that this quantity acts as an entanglement witness based on a well-known test closely related to the entropic and reduction criteria of separability~\cite{HorodeckiEntropy1996,NielsenKempe2001,HorodeckiReduction1999,Cerf1999}. The relation between purities and entanglement has been further explored in recent works~\cite{SciRep2023,npjQI2024}, including optical experiments in which purities are measured directly~\cite{Bovino2005,Walborn2006,npjQI2024}. The main contribution of this work is to show that the four HOM measurements already employed in Ref.~\cite{Temporao2024} give direct access to both purities; in particular, to the purity of the non-polarization DOFs, without experimentally resolving them. Even though the witness is linear in the coincidence probabilities, it is quadratic in the density operator, resulting in a nonlinear, two-copy entanglement witness, rather than a conventional linear functional of the density operator~\cite{GuhneToth2009}.

The witness is applied to a realistic model of quantum-dot photons, in which it provides a sufficient condition for verifying polarization--frequency entanglement induced by the fine-structure splitting in the presence of dephasing, without resolving the photon in time or frequency. We compare it with a complementary criterion that combines global purity with the conventionally measured DOP and identify the experimental conditions under which a positive value is a reliable certificate. Section~\ref{sec:method} introduces HOM-based mode-resolved Stokes operators, Sec.~\ref{sec:witness} introduces the witness, and Sec.~\ref{sec:examples} applies it to examples, including quantum-dot photons. Numerical tests are reported in Sec.~\ref{sec:sim}, experimental imperfections are discussed in Sec.~\ref{sec:imperf}, and generalizations and limitations in Sec.~\ref{sec:ext}; all proofs are given in the appendices.

%=====================================================================
\section{Method}\label{sec:method}
%=====================================================================
Let $\mathcal H=\Hp\otimes\Ho$ be the Hilbert space of a photon, where $\Hp$ is the (two-dimensional) subspace associated with polarization and $\Ho$ is associated with all the remaining internal DOFs of the photon, such as time of arrival, frequency, spatial mode and so on. Any other quantum system external to the photon belongs to the environment $E$, represented by Hilbert space $\mathcal \He$. 

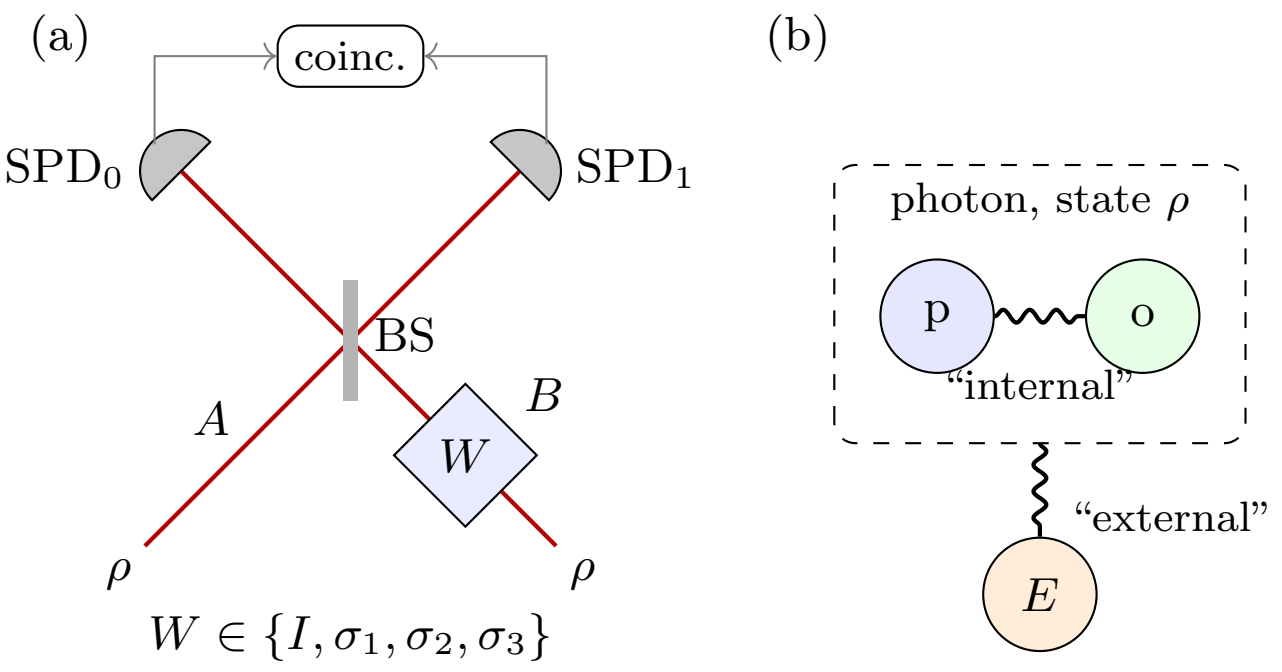
\begin{figure}[t]
\centering
\begin{tikzpicture}[scale=0.8,every node/.style={font=\small}]
  % beams
  \draw[thick,red!70!black] (-1.7,-1.7) -- (0,0) -- (1.4,1.4);
  \draw[thick,red!70!black] (1.7,-1.7) -- (0,0) -- (-1.4,1.4);
  % beam splitter (vertical, as in Ref. [Temporao2024])
  \fill[gray!60] (-0.06,-0.5) rectangle (0.06,0.5);
  \node at (0.45,0.05) {BS};
  % W operation
  \node[draw,fill=blue!8,diamond,minimum size=0.95cm,inner sep=0pt] at (0.95,-0.95) {$W$};
  % inputs
  \node[below left=-1pt] at (-1.7,-1.7) {$\rho$};
  \node[below right=-1pt] at (1.7,-1.7) {$\rho$};
  \node at (-1.15,-0.65) {$A$};
  \node at (1.6,-0.45) {$B$};
  % detectors: flat face perpendicular to the beam
  \draw[fill=gray!45] (-1.4,1.4) ++(45:0.34) arc[start angle=45,end angle=225,radius=0.34] -- cycle;
  \draw[fill=gray!45] (1.4,1.4) ++(-45:0.34) arc[start angle=-45,end angle=135,radius=0.34] -- cycle;
  \node[anchor=east] at (-1.72,1.42) {SPD$_0$};
  \node[anchor=west] at (1.72,1.42) {SPD$_1$};
  \node[draw,rounded corners,fill=white,font=\footnotesize] (C) at (0,2.35) {coinc.};
  \draw[->,gray] (-1.62,1.62) |- (C.west);
  \draw[->,gray] (1.62,1.62) |- (C.east);
  \node at (0,-2.45) {$W\in\{I,\sigma_1,\sigma_2,\sigma_3\}$};
  % (b) photon Hilbert space
  \begin{scope}[xshift=5.7cm,yshift=0.35cm]
    \draw[dashed,rounded corners] (-1.7,-1.2) rectangle (1.7,1.1);
    \node at (0,0.8) {\footnotesize photon, state $\rho$};
    \node[circle,draw,fill=blue!10,minimum size=0.75cm] (p) at (-0.85,-0.15) {p};
    \node[circle,draw,fill=green!10,minimum size=0.75cm] (o) at (0.85,-0.15) {o};
    \draw[decorate,decoration={snake,amplitude=1.2pt,segment length=5pt},thick] (p) -- (o);
    \node[below] at (0,-0.45) {\footnotesize ``internal''};
    \node[circle,draw,fill=orange!15,minimum size=0.75cm] (E) at (0,-2.45) {$E$};
    \draw[decorate,decoration={snake,amplitude=1.2pt,segment length=5pt},thick] (0,-1.2) -- (E);
    \node[right] at (0.1,-1.8) {\footnotesize ``external''};
  \end{scope}
  \node at (-2.4,2.5) {(a)};
  \node at (3.7,2.5) {(b)};
\end{tikzpicture}
\caption{(a) Scheme of the method of Ref.~\cite{Temporao2024}. Two photons in state $\rho$ interfere in a beam splitter (BS) after one of them is subjected to the unitary operation $W$ acting on $\mathcal \Hp$, and the probability of coincidence counts between the single-photon detectors SPD$_0$ and SPD$_1$ is measured for the four settings $W\in\{I,\sigma_1,\sigma_2,\sigma_3\}$. No polarization-selective component is employed. (b) The polarization (p) of a photon may be entangled with its other internal DOFs (o), with an external environment ($E$), or with both. The witness introduced in Sec.~\ref{sec:witness} detects the former even in the presence of the latter.}
\label{fig:setup}
\end{figure}

Figure~\ref{fig:setup}(a) shows the basic setup, which is identical to the one of Ref.~\cite{Temporao2024}: two identical photons, described by the density operator $\rho=\tr_E\ket\Psi\!\bra\Psi$ obtained by tracing out the environment from a global (pure) state $\ket\Psi \in \mathcal H \otimes \He$. They interfere in a 50/50 beamsplitter (BS) after one of them undergoes a unitary operation $W$ acting on its polarization DOF. We then measure the probability of coincidence counts between the single-photon detectors (SPDs) placed at the BS outputs while the operation $W$ is sequentially chosen among the identity $I$ and the three Pauli operators $\sigma_1$, $\sigma_2$, and $\sigma_3$, which, for polarization, can be implemented by birefringent elements.

Now let $\rhop\equiv\tr_{\mathrm o}\rho$ and $\rhoo\equiv\tr_{\mathrm p}\rho$ be the reduced density operators describing, respectively, the polarization and the remaining DOFs. Standard polarimetry (QST) only has access to $\rhop$, i.e., to the Stokes parameters averaged over all the other DOFs, $\langle s_j\rangle=\tr(\rhop\sigma_j)$, $j=1,2,3$, which we call \emph{mode-averaged Stokes parameters}; from then, the $\mathrm{DOP}=(\sum_j\langle s_j\rangle^2)^{1/2}$ can be computed. Note that the Pauli operators in this work are (unusually) labeled to match the Stokes parameters, i.e., $\sigma_1 \equiv Z$, $\sigma_2 \equiv X$, and $\sigma_3 \equiv Y$ correspond to the horizontal/vertical, diagonal/antidiagonal, and right-/left-circular polarization bases, respectively, and we also define $\sigma_0\equiv I$. Throughout this work we assume that the operations $W$ act only on the polarization Hilbert space, i.e., that each setting implements $\sigma_j\otimes\Io$, where $\Io$ is the identity on $\mathcal \Ho$. This assumption is not as trivial as it may seem, as the other DOFs are precisely those we wish to characterize; the consequences of relaxing this assumption are discussed in Secs.~\ref{sec:optics} and \ref{sec:pdl}.

\subsection{``Internal'' and ``external'' entanglement}

We say that the photon presents \emph{internal entanglement} whenever its polarization state is entangled with at least one of its other internal DOFs, i.e., when $\rho$ cannot be written as a statistical mixture of states of the form $\alpha_k\otimes\beta_k$, with $\alpha_k$ acting on $\Hp$ and $\beta_k$ acting on $\Ho$~\cite{HorodeckiRMP2009,GuhneToth2009}. If only internal entanglement is present, $\rho$ is a pure state. Entanglement with the environment, however, renders $\rho$ mixed; we define such states as having \emph{external entanglement}. Mixed states can also be constructed from incoherent mixtures, in which the source emits pure states according to a certain probability distribution; for measurements on single photons, these two mechanisms are indistinguishable, but in the current scheme they have very distinct effects; see Sec. \ref{sec:drift}. In general, a photon may simultaneously present both kinds of entanglement [Fig.~\ref{fig:setup}(b)].

\subsection{Two-photon interference with mixed states}

Let $\mathcal{F}(U)$ be the fidelity between the two input states when the unitary operator $U$ is applied to the polarization of one of them. It can be shown~\cite{Temporao2024} that the coincidence probability, as a function of $U$, is given by $\Pc(U)=[1-\mathcal{F}(U)]/2$. The same expression holds for mixed states (Appendix~\ref{app:hom}), provided the fidelity is defined as the overlap between the density operators of the two photons:
\begin{equation}
    \label{eq:F}
    \mathcal{F}(U)=\tr\bigl[\rho\,(U\otimes\Io)\,\rho\,(U\otimes\Io)^\dagger\bigr],
\end{equation}
where $\Io$ is the identity operator acting on $\Ho$ as before. $\mathcal{F}(U)$ measures the similarity between the two photons when they reach the BS: identical pure photons always bunch ($\mathcal{F}=1$), whereas orthogonal photons follow random trajectories ($\mathcal{F}=0$). For the sake of brevity, from now on define $P_j\equiv\Pc(\sigma_j)$ and $\mathcal{F}_j\equiv \mathcal{F}(\sigma_j)$, for $j=0,\dots,3$.

The first useful result comes from the identity setting. When $U=I$, Eq.~\eqref{eq:F} gives $\mathcal{F}_0=\tr\rho^2$, the purity of the photon, so that
\begin{equation}
\tr\rho^2=1-2P_0 ,
\label{eq:purity}
\end{equation}
which is the foundation of a well-known purity test: $P_0=0$ if, and only if, the two copies of the photon are in the same pure state.\footnote{In Sec.~III\,C of Ref.~\cite{Temporao2024} this relation appears as $P_{\mathrm{coinc}}(I)=\tr\rho^2$ owing to a misprint; Eq.~\eqref{eq:purity} is the form consistent with Eqs.~(2) and (13) therein.}

\subsection{Mode-resolved Stokes operators}

In Ref.~\cite{Temporao2024}, the settings $\{\sigma_i\}_{i=0}^4$ were shown to yield the squared Stokes parameters in two particular cases: pure states, and states where the polarization is entangled with the environment but whose other DOFs are in a pure state uncorrelated with the polarization, $\rho=\rhop\otimes\ket\xi\!\bra\xi_\text{o}$. In the latter case, the other DOFs factor out of the HOM fidelities, and the photon is solely described by its polarization. In both cases $\langle s_j\rangle^2=1-2P_j-2P_0$, which simplifies to $1-2P_j$ for pure states, since then $P_0=0$~\cite{Temporao2024}. Now, for the general case, consider the state
\begin{equation}
\ket\Psi=\alpha\ket{H}\ket{t_0}+\beta\ket{V}\ket{t_1} , 
\label{eq:entangled}
\end{equation}
where the time bins $t_0$ and $t_1$ are such that $|t_0-t_1|$ is much longer than the photon coherence time. A slow detector, unable to resolve the two time bins, sees a partially polarized photon, with $\langle s_1\rangle=|\alpha|^2-|\beta|^2$ and $\langle s_2\rangle=\langle s_3\rangle=0$~\cite{Temporao2024}. A detector fast enough to resolve them, on the other hand, would see perfectly polarized states in each bin: horizontal at $t_0$ and vertical at $t_1$. The DOP obtained by a slow polarimer is reduced not because the polarization varies randomly, but because it \emph{is not constant within the photon}.

In order to capture this structure, we define the \emph{mode-resolved Stokes operators}
\begin{equation}
A_j\equiv\tr_{\mathrm p}\bigl[(\sigma_j\otimes\Io)\rho\bigr],\qquad j=0,\dots,3,
\label{eq:A}
\end{equation}
which act on $\Ho$. Note that $A_0=\rhoo$. If $\Ho$ describes the time of arrival DOF, the diagonal elements of $A_j$ are the time-resolved Stokes parameters, while its off-diagonal elements describe coherences between different times. Note that the mode-averaged Stokes parameters are simply their traces, $\langle s_j\rangle=\tr A_j$. For the state of Eq.~\eqref{eq:entangled}, for instance,
\begin{align}
A_0&=|\alpha|^2\ket{t_0}\!\bra{t_0}+|\beta|^2\ket{t_1}\!\bra{t_1},\nonumber\\
A_1&=|\alpha|^2\ket{t_0}\!\bra{t_0}-|\beta|^2\ket{t_1}\!\bra{t_1},\nonumber\\
A_2&=\alpha\beta^*\ket{t_0}\!\bra{t_1}+\alpha^*\beta\ket{t_1}\!\bra{t_0},\nonumber\\
A_3&=i\bigl(\alpha\beta^*\ket{t_0}\!\bra{t_1}-\alpha^*\beta\ket{t_1}\!\bra{t_0}\bigr).\nonumber
\end{align}
The operators $A_0$ and $A_1$ are diagonal in the time-bin basis, their traces recovering $\langle s_0\rangle=1$ and $\langle s_1\rangle=|\alpha|^2-|\beta|^2$. The operators $A_2$ and $A_3$, on the other hand, are purely off-diagonal: not only their traces, $\langle s_2\rangle$ and $\langle s_3\rangle$, vanish, but also their diagonal elements, so that even time-resolved polarimetry finds no diagonal or circular polarization component, as evident from Eq. \ref{eq:entangled}.

We show in Appendix~\ref{app:stokes} that each of the four coincidence probabilities depends only on the four quantities $\norm{A_j}^2\equiv\tr A_j^2$. In other words, \emph{standard polarimetry measures $\tr A_j$, whereas HOM interference measures $\tr A_j^2$}. There is an immediate analogy with the difference between the squared mean $\langle s\rangle^2$ and the mean square $\langle s^2\rangle$ of a signal: the latter may remain nonzero even when the signal averages out to zero. In the example above, $\langle s_2\rangle=0$, but $\norm{A_2}^2=2|\alpha\beta|^2\neq0$, since the coherence between the two time bins still exists. This is exactly the kind of information that slow polarimetry discards and two-photon interference retains. If the photons are not perfectly indistinguishable in the other DOFs but polarization is not coupled to them, all four fidelities $\mathcal{F}_j$ are simply multiplied by a factor $\tr\rhoo^2$, which corresponds to the usual HOM visibility.

\subsection{Erasing the polarization}

Applying the unitaries $I$, $\sigma_1$, $\sigma_2$, and $\sigma_3$ with equal probabilities to one of the photons, and averaging out the four HOM coincidence probabilities, implements an operation known as the Pauli twirl~\cite{Bennett1996}. It acts as a completely depolarizing channel on the polarization DOF while leaving the reduced state of all the other DOFs unchanged (Appendix~\ref{app:stokes}):
\begin{equation}
\frac14\sum_{j=0}^3(\sigma_j\otimes\Io)\,\rho\,(\sigma_j\otimes\Io)=\frac{\Ip}{2}\otimes\rhoo ,
\label{eq:twirlrho}
\end{equation}
where $\Ip$ is the identity acting on $\Hp$; the photon is thus replaced by a product state, in which the polarization is randomized and uncorrelated with the other DOFs.  The polarization is thus ``erased'' from the interference, which now only tests whether the \emph{remaining} DOFs of the two photons are identical; in other words, it becomes a ST restricted to the non-polarization DOFs. Indeed, averaging the four fidelities of Eq.~\eqref{eq:F} and using Eq.~\eqref{eq:twirlrho} yields
\begin{equation}
\frac14\sum_{j=0}^3\mathcal{F}_j=\tr\Bigl[\rho\Bigl(\frac{\Ip}{2}\otimes\rhoo\Bigr)\Bigr]=\frac12\tr\rhoo^2 ,
\label{eq:avgF}
\end{equation}
or, in terms of the coincidence probabilities,
\begin{equation}
\tr\rhoo^2=2-(P_0+P_1+P_2+P_3).
\label{eq:sumrule}
\end{equation}

Equations~\eqref{eq:purity} and~\eqref{eq:sumrule}, when taken together, mean that the four Pauli measurements give access to the purity of the photon both \emph{as a whole} and \emph{without its polarization}. As we shall see next, by comparing these two numbers one is able to detect internal entanglement.

%=====================================================================
\section{The entanglement witness}\label{sec:witness}
%=====================================================================

\subsection{Main result}

The main idea of the entanglement witness comes from the fact that if a quantum state $\rho$ is separable, then it is at most as pure as any one of its subsystems~\cite{HorodeckiEntropy1996,NielsenKempe2001,HorodeckiReduction1999,Cerf1999}. Employing the notation of our present case, if the photon has no internal entanglement, then
\begin{equation}
\tr\rho^2\le\tr\rhoo^2 .
\label{eq:ineq}
\end{equation}
See Appendix~\ref{app:proof} for more details. Combining Eq.~\eqref{eq:ineq} with Eqs.~\eqref{eq:purity} and \eqref{eq:sumrule}, we arrive at the main result of this work: the \emph{entanglement witness} defined by
\begin{equation}
\W\equiv P_1+P_2+P_3-P_0-1=\tr\rho^2-\tr\rhoo^2
\label{eq:W}
\end{equation}
never takes positive values for photons without internal entanglement. Therefore, an experimental value $\W>0$ certifies that the polarization is entangled with another internal DOF of the photon, \emph{irrespective of how much entanglement with the environment, or classical mixing, is also present}. The novelty here is not the criterion of Eq.~\eqref{eq:ineq}, but the direct access to it provided by the four coincidence probabilities of Ref.~\cite{Temporao2024}, with no additional measurements and without resolving the other DOFs.

Note that, in terms of the R\'enyi-2, or collision, entropy $S_2(\rho)=-\log_2\tr\rho^2$~\cite{HorodeckiEntropy1996}, Eq.~\eqref{eq:ineq} reads $S_2(\rho)\ge S_2(\rhoo)$. A positive witness means, therefore, that the photon has \emph{less} entropy as a whole than its non-polarization part, which is impossible without entanglement. For pure states, $S_2(\rhoo)$ is the R\'enyi-2 entanglement entropy between polarization and the other DOFs, which ranges from 0 to 1.

Additional information can be obtained by combining $\W$ with the purity test of Eq.~\eqref{eq:purity}, i.e., with $P_0$. Four possible scenarios arise, which are summarized in Table~\ref{tab:class}. If $P_0=0$, the photon is in a pure state; it immediately follows that $\W\ge0$, with equality for fully polarized photons and $\W>0$ for internally entangled ones. If $P_0>0$, the photon is in a mixed state, which
may result from external entanglement or classical mixing. A positive $\W$ nevertheless certifies internal entanglement, whereas $\W\le0$ means that internal entanglement detection has failed (see Sec.~\ref{sec:limits}).

\begin{table}[t]
\caption{Classification of single photons according to the coincidence probability with the identity setting, $P_0$, and to the witness $\W$ of Eq.~\eqref{eq:W}.}
\label{tab:class}
\begin{ruledtabular}
\begin{tabular}{ccl}
$P_0$ & $\W$ & Conclusion\\
\hline
$=0$ & $=0$ & pure product state (fully polarized photon)\\
$=0$ & $>0$ & pure, internally entangled~\cite{Temporao2024}\\
$>0$ & $>0$ & mixed \emph{and} internally entangled\\
$>0$ & $\le0$ & mixed; internal entanglement not detected\\
\end{tabular}
\end{ruledtabular}
\end{table}

\subsection{A complementary witness}
The reduction criterion for separability also implies that, for every photon without internal entanglement, the photon as a whole is at most as pure as its \emph{polarization} part, $\tr\rho^2\le\tr\rhop^2$. Since polarization is a qubit, $\tr\rhop^2=(1+\mathrm{DOP}^2)/2$, thus the quantity
\begin{equation}
\W_{\mathrm p}\equiv\tr\rho^2-\frac{1+\mathrm{DOP}^2}{2}=\frac{1-4P_0-\mathrm{DOP}^2}{2}
\label{eq:Wp}
\end{equation}
also acts as an entanglement witness. The two criteria compare the photon with each of its two parts and, as shown in Sec.~\ref{sec:qd}, each of them can detect internal entanglement that the other can't, so they can be used together. Note, however, that a photon that is both mixed and internally entangled cannot have its DOP extracted from the four HOM measurements alone (Sec.~\ref{sec:limits}) and must be measured by conventional polarimetry, i.e., with polarization-selective components, a clear disadvantage when compared to $\W$.

\subsection{Features of the witness}

Some features of $\W$ are worth highlighting. First, for pure states the witness is both necessary and sufficient: a pure photon is internally entangled if, and only if, $\W>0$. In this case, there is a direct connection to the DOP via $\W=(1-\mathrm{DOP}^2)/2$, so the witness quantifies how much of the depolarization is due to internal entanglement. Also note that, in this case, $\W$ and $\W_{\mathrm p}$ convey the same information.

Second, no alignment of the measurement axes is required. The depolarizing effect of the four settings does not depend on the absolute orientation of the wave plates, as long as the eigenstates of $\sigma_1$, $\sigma_2$, and $\sigma_3$ correspond to three mutually orthogonal axes of the Poincaré sphere. In other words, $\W$ is rotationally invariant.

Third, the witness is bounded. For two independent photons in the same state $\rho$, it is possible to show that $-\frac12\le\W\le\frac12$ (see Appendix~\ref{app:proof}). The maximum is reached by maximally entangled pure states, such as the state of Eq.~\eqref{eq:entangled} with $|\alpha|=|\beta| = 1/\sqrt{2}$. The minimum is reached by a completely unpolarized photon whose other DOFs are in a pure state, i.e., by purely external depolarization. The range is wider when these assumptions are relaxed, as also shown in Appendix~\ref{app:proof}. 

Finally, it can be shown that a set of four unitary operations is the minimum required to completely depolarize a qubit~\cite{Gross2007}. There is also a simple geometric picture behind the witness: the norms $\norm{A_j}$, $j=1,2,3$, can be seen as the components of a ``mode-resolved Stokes vector'', and $\norm{A_0}$ as the corresponding intensity. Without internal entanglement, this vector is always shorter than the intensity, which is the analog of the familiar condition $\mathrm{DOP}\le1$. Internal entanglement, however, can make it up to $\sqrt3$ times longer (Appendix~\ref{app:proof}), even when the mode-averaged Stokes vector approaches zero.

%=====================================================================
\section{Examples}\label{sec:examples}
%=====================================================================

\subsection{Distinguishing between ``internal'' and ``external'' entanglement}

Let us first apply the witness to two simple scenarios, starting with the case of Eq.~\eqref{eq:entangled}. The photon, considered as a whole, is a pure state ($\tr\rho^2=1$), whereas its time-bin part is mixed, with $\tr\rhoo^2=(1+\mathrm{DOP}^2)/2$. Equation~\eqref{eq:W} then gives $\W=(1-\mathrm{DOP}^2)/2>0$ whenever $\mathrm{DOP}<1$. Now, for a photon whose polarization is entangled only with the environment, $\tr\rho^2=(1+\mathrm{DOP}^2)/2$, while its other DOFs are in a pure state, $\tr\rhoo^2=1$. Therefore, $\W=-(1-\mathrm{DOP}^2)/2\le0$. Note that both mechanisms lead to the same DOP and are indistinguishable by standard polarimetry, but the witness places them symmetrically around zero.

A third situation of practical interest is that of photons that are not perfectly indistinguishable in their other DOFs, e.g., due to spectral impurity, while their polarization is not coupled to those DOFs. As mentioned in Sec.~\ref{sec:method}, all four fidelities are then multiplied by the same factor $\tr\rhoo^2<1$, and one finds $\W=\tr\rhoo^2(\tr\rhop^2-1)\le0$. This means that no distinguishability of the photons, due to imperfections of real sources, is able to produce false positives.

\subsection{Both mechanisms at once}\label{sec:mixed}

\begin{figure}[t]
\centering
\includegraphics[width=\columnwidth]{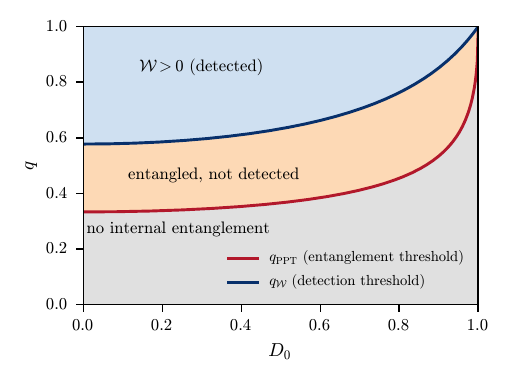}
\caption{Internal entanglement of the photon described by Eq.~\eqref{eq:model}, as a function of the DOP $D_0$ of the internally entangled component and of its weight $q$. Gray: no internal entanglement. Orange: internally entangled, but not detected by the witness. Blue: internally entangled and detected, $\W>0$.}
\label{fig:regions}
\end{figure}

Let us now consider a photon presenting both kinds of entanglement. A simple model is a photon that is prepared, with probability $q$, in the pure state $\ket\Psi$ of Eq. \ref{eq:entangled}, and in a completely mixed state of polarization and time bin with probability $1-q$:
\begin{equation}
\rho(q)=q\ket\Psi\!\bra\Psi+(1-q)\,\frac{I_4}{4},
\label{eq:model}
\end{equation}
where $I_4$ is the identity operator on the four-dimensional Hilbert space spanned by polarization and time-bin. We choose $|\alpha|^2=(1+D_0)/2$ and $|\beta|^2=(1-D_0)/2$, so that $D_0$ is the DOP of $\ket\Psi$, which is internally entangled for $D_0<1$; the DOP of the photon is then $qD_0$. It is straightforward to show that the coincidence probabilities are given by
\begin{equation}
\begin{aligned}
P_0&=\tfrac38(1-q^2),\qquad P_2=P_3=\tfrac18(3+q^2),\\
P_1&=\tfrac18(3+q^2-4q^2D_0^2),
\end{aligned}
\label{eq:Pmodel}
\end{equation}
and therefore
\begin{equation}
\W=\tfrac14\bigl[q^2(3-2D_0^2)-1\bigr].
\label{eq:Wmodel}
\end{equation}
The purity test detects mixedness ($P_0>0$) whenever $q<1$, but cannot distinguish external entanglement from classical mixing, nor reveal internal entanglement that may
coexist with either mechanism. The witness, on the other hand, detects internal entanglement whenever
$q>q_W=(3-2D_0^2)^{-1/2}$. For comparison, a complete analysis of this state, based on the partial-transpose~\cite{Peres1996,Horodecki1996PLA}, shows that it is entangled whenever $q>q_{\mathrm{PPT}}=[1+2(1-D_0^2)^{1/2}]^{-1}$. Indeed, the partial transpose of $\rho(q)$ with respect to polarization has, in the subspace spanned by $\ket H\ket{t_1}$ and $\ket V\ket{t_0}$, the eigenvalues $\frac14(1-q)\pm q|\alpha\beta|$, the smaller of which is negative whenever $q>q_{\mathrm{PPT}}$, whereas its other two eigenvalues are always positive; for two qubits, a positive partial transpose also implies separability~\cite{Horodecki1996PLA}. These two thresholds are shown in Fig.~\ref{fig:regions}.

Note that when $D_0=0$ the internal DOFs are maximally entangled, and the witness detects this entanglement for $q>1/\sqrt3\approx0.58$, whereas the state is entangled for $q>1/3$. In the gap between these values (orange region in Fig.~\ref{fig:regions}), internal entanglement is present but too weak to be detected by this scheme; this is the price paid for using only two numbers (the two purities) instead of a full QST, which would require resolving the time bins. Second, as $D_0\to1$ both thresholds approach $q=1$: the amplitude $|\beta|$ vanishes, and $\ket\Psi$ becomes $\ket H\ket{t_0}$, a product state, so that no internal entanglement is left, and $\rho(q)$ is separable for every $q$.

\subsection{Application to quantum-dot photons}\label{sec:qd}

A physically motivated example in which both internal and external depolarization may arise is provided by photons emitted by a semiconductor quantum dot (QD)~\cite{Santori2002,Stevenson2006,Akopian2006}. The two bright states of the neutral exciton, split in energy by the fine-structure splitting (FSS) $S \equiv \hbar \Omega$, where $\Omega$ is the corresponding angular frequency, decay by emitting horizontally and vertically polarized photons, respectively~\cite{Stevenson2006,Akopian2006}. If the exciton is resonantly excited with a polarization at an angle $\theta$ with respect to the FSS axes, the emitted photon is described by
\begin{equation}
\ket\psi=\cos\theta\ket H\ket{\varphi_+}+\sin\theta\ket V\ket{\varphi_-},
\label{eq:qd}
\end{equation}
where $\ket{\varphi_\pm}$ are exponentially decaying wave packets with radiative decay rate $\Gamma$ and central angular frequencies $\omega_0\pm \Omega/2$. In the time domain, the polarization of this photon precesses at the angular frequency $\Omega$ during the emission, as its polarization and frequency are entangled to each other.

Here we include two decoherence processes (see Appendix~\ref{app:qdmodel}): pure dephasing at a rate $\gamma^*$, caused, e.g., by phonons, which affects both transitions in the same way and reduces the HOM visibility to $M=\Gamma/(\Gamma+2\gamma^*)$; and a coherence reduction factor $\eta\le1$ between the two bright states, which represents any process that entangles the polarization with the environment. The case $\eta=0$ describes, in particular, the unconditional reduced state of the exciton photon emitted in an ideal biexciton–exciton cascade~\cite{Stevenson2006,Akopian2006}, whose polarization is entangled with that of the biexciton photon. In this case, the exciton photon considered alone is in a statistical mixture of $\ket H\ket{\varphi_+}$ and $\ket V\ket{\varphi_-}$, in which polarization and frequency are classically correlated, but not entangled. We assume that photons emitted in successive excitation cycles are independently and identically prepared. Even though the exciton returns to the ground state after each emission, correlations between successive photons may arise from slowly varying environmental degrees of freedom.

\begin{figure}[t]
\centering
\includegraphics[width=\columnwidth]{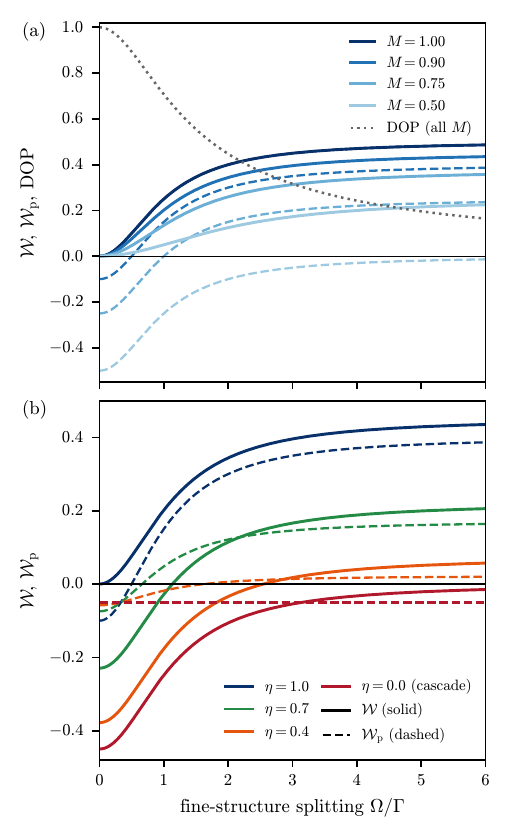}
\caption{Witness for photons emitted by a QD exciton excited with diagonal polarization ($\theta=45^\circ$), as a function of the fine-structure splitting $\Omega = S/\hbar$ (in angular-frequency units) normalized to the radiative decay rate $\Gamma$, for independent photons. Solid lines: witness $\W$, Eq.~\eqref{eq:Wqd}; dashed lines: complementary criterion $\W_{\mathrm p}$, Eq.~\eqref{eq:Wp}. (a) Perfect coherence between the bright states ($\eta=1$) and several values of the HOM visibility $M=\Gamma/(\Gamma+2\gamma^*)$ of each polarization component; the dotted line shows the time-integrated DOP, which does not depend on $M$. (b) $M=0.9$ and several values of the coherence factor $\eta$; $\eta=0$ corresponds to the unconditional exciton-photon state in an ideal biexciton--exciton cascade.}
\label{fig:qd}
\end{figure}

For this model, the witness can be calculated in closed form (Appendix~\ref{app:qdmodel}):
\begin{gather}
\W=\frac M2\sin^2(2\theta)\Bigl[\eta^2-\frac{1}{1+\tilde \Omega^2}\Bigr],\label{eq:Wqd}\\
\tilde \Omega\equiv\frac{\Omega}{\Gamma+2\gamma^*},\nonumber
\end{gather}
where $\Gamma+2\gamma^*$ is the homogeneous spectral
linewidth (FWHM, in angular-frequency units) of each
polarization resolved optical transition. Thus,
$\tilde\Omega$ measures the fine-structure splitting
relative to the linewidth. 

As the excitation angle enters Eq.~\eqref{eq:Wqd}
only as a scaling factor, we take $\theta=45^\circ$
in Fig.~\ref{fig:qd}. For $\eta=1$
[Fig.~\ref{fig:qd}(a)], $\W$ is positive for
every $\Omega>0$, regardless of pure dephasing,
which reduces its magnitude but not its sign.
For $\eta<1$ [Fig.~\ref{fig:qd}(b)], a detection
threshold appears:
$\tilde\Omega>\sqrt{1-\eta^2}/\eta$.
For example, at $M=0.9$ and $\eta=0.7$,
entanglement is detected only for
$\Omega/\Gamma>1.13$. In the cascade limit
$\eta=0$, $\W$ is never positive, even though
polarization and frequency may be strongly
correlated. The splitting and linewidth ranges
shown are representative of self-assembled
quantum dots \cite{Prechtel2013}.

The dashed curves in Fig.~\ref{fig:qd}
correspond to the complementary criterion
$\W_{\mathrm p}$ of Eq.~\eqref{eq:Wp}, obtained
from the global purity and the time-integrated
degree of polarization (DOP) -- see Appendix~\ref{app:qdmodel}. The two witnesses
exhibit complementary sensitivities. For
$M=0.9$ and $\eta=0.7$, $\W_{\mathrm p}$
detects entanglement already at
$\Omega/\Gamma>0.66$, compared with $1.13$
for $\W$. Conversely, $\W$ is more robust
against pure dephasing: for $\eta=1$ it
remains positive for every $M>0$ and
$\Omega>0$, whereas $\W_{\mathrm p}$ cannot
detect entanglement when $M\leq1/2$.
The criteria can thus be combined through
$\max(\W,\W_{\mathrm p})>0$.

More fundamentally, the DOP and global purity
do not uniquely determine internal entanglement.
For instance, the states with
$(\eta,\Omega/\Gamma,M)=(1,\sqrt3,0.6)$
and $(0.5,0,0.96)$ have identical DOP
of $0.5$ and global purity of $0.6$, but
yield \(\mathcal W\simeq0.16\) and \(\mathcal W\simeq-0.36\),
respectively. Only the former is internally
entangled. This example shows that the HOM witness provides
information not contained in conventional
characterizations.

Even though the expected value of $\W$ can be
predicted from source parameters, note that its
experimental certification is independent of
the source model. Time- or frequency-resolved
polarimetry~\cite{Stevenson2008,Njala2026}
reveals polarization--frequency correlations,
but cannot by itself distinguish entangled
states from separable states with the same
resolved polarization statistics
(Appendix~\ref{app:qdmodel}).
A positive witness, on the other hand, is able to certify
entanglement without temporal or spectral
resolution, subject to the experimental
conditions of Sec.~\ref{sec:imperf}.

%=====================================================================
\section{Numerical tests}\label{sec:sim}
%=====================================================================

Using the \textsc{qiskit} framework~\cite{Qiskit} and the equivalence between the HOM effect and the swap test~\cite{GarciaEscartin2013,Ripper2023,Foulds2021}, we checked our analytical predictions: each DOF of a photon is represented by a qubit, the environment by qubits that are traced out, and the coincidence probability by the parity of two swap-test ancillas (Appendix~\ref{app:circuit}). For the model of Sec.~\ref{sec:mixed}, the simulated witness agrees with Eq.~\eqref{eq:Wmodel} within the statistical fluctuations and changes sign at $q_{\W}$, and the exact output probabilities of all 336 circuits agree with a direct density-matrix calculation to within $10^{-15}$ (Fig.~\ref{fig:sim} -- see also Fig.~\ref{fig:circuit} in Appendix~\ref{app:circuit}).

\subsection{Witness sensitivity to mixed-state entanglement}

To assess the detection capability beyond the
specific family of states considered above, we
generated 10\,000 random polarization--time-bin
density operators, with ranks from 1 to 4.
For pure states, the analytical relation
$\W=2\mathcal N^2$ ensures the detection of
every entangled state, where $\mathcal N$ is
the negativity. Among the sampled entangled
mixed states, the fractions detected were
49\%, 17\%, and 6\% for ranks 2, 3, and 4,
respectively. These values characterize the
sampled ensembles and illustrate the limited
sensitivity of the criterion to mixed-state
entanglement. No false positives were observed.

\begin{figure}[t]
\centering
\includegraphics[width=\columnwidth]{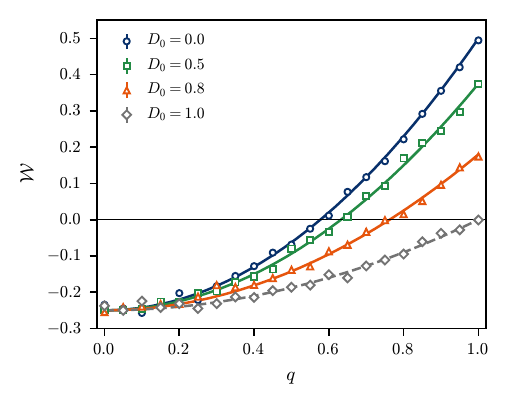}
\caption{Witness $\W$ as a function of $q$ for the state of Eq.~\eqref{eq:model}, for several values of $D_0$. Lines: theoretical values given by Eq.~\eqref{eq:Wmodel}. Symbols: simulations using the circuit of Fig.~\ref{fig:circuit} (Appendix~\ref{app:circuit}), with $10^4$ shots per Pauli setting. The dashed line ($D_0=1$) corresponds to a photon without internal entanglement.}
\label{fig:sim}
\end{figure}

\subsection{Statistical uncertainty}\label{sec:stat}

The number of photon pairs needed in an actual experiment is set by the statistical uncertainty of the witness. Each coincidence probability is estimated from a finite number $N$ of photon pairs; since each pair either produces a coincidence or not, the estimate of $P_j$ follows a binomial distribution, with standard deviation $[P_j(1-P_j)/N]^{1/2}\le1/(2\sqrt N)$. As $\W$ is the sum of four independent estimates, its standard deviation is
\begin{equation}
\sigma_{\W}=\Bigl[\sum_{j=0}^3\frac{P_j(1-P_j)}{N}\Bigr]^{1/2}\le\frac{1}{\sqrt N},
\label{eq:var}
\end{equation}
i.e., about $0.009$ for the $N=10^4$ shots used in the simulations. For the model of Sec.~\ref{sec:mixed}, $\sigma_{\W}\sqrt N$ lies between 0.7 and 1. Hence, in order to claim a violation with a confidence of $Z$ standard deviations, one needs $N\gtrsim Z^2/\W^2$ photon pairs per setting, a number that grows quadratically as the state approaches the detection threshold (Table~\ref{tab:stats}). Even at a modest rate of $10^3$ detected pairs per second, these numbers correspond to measurement times ranging from a fraction of a second to about one minute per setting. Note that these are idealized estimates: they do not include the additional pairs needed for calibration described in Sec.~\ref{sec:bs}, whose own statistical uncertainty adds to that of Eq.~\eqref{eq:var}, nor systematic uncertainties or setting-dependent losses.

\begin{table}[t]
\caption{Witness value and number $N$ of photon pairs per Pauli setting required for a violation by five standard deviations, for the state of Eq.~\eqref{eq:model} with $D_0=0$.}
\label{tab:stats}
\begin{ruledtabular}
\begin{tabular}{ccccccc}
$q$ & 0.60 & 0.65 & 0.70 & 0.80 & 0.90 & 1.00\\
\hline
$\W$ & 0.020 & 0.067 & 0.118 & 0.230 & 0.358 & 0.500\\
$N$ & $5.7\times10^4$ & $5.1\times10^3$ & $1.6\times10^3$ & 407 & 160 & 75\\
\end{tabular}
\end{ruledtabular}
\end{table}

%=====================================================================
\section{Experimental imperfections}\label{sec:imperf}
%=====================================================================

The previous sections assumed ideal experimental conditions. A positive value of $\W$ reliably implies internal entanglement only if the following requirements are met: (i) the two photons must be independent; classically correlated photons can also be used, but the certificate then refers to their conditional states (Sec.~\ref{sec:nonid}); (ii) the optical components must act on the polarization alone (Sec.~\ref{sec:optics}), and no polarization-dependent loss may exist after them (Sec.~\ref{sec:pdl}); (iii) the coincidence counts must be correctly converted into the fidelities $\mathcal{F}_j$, which is best achieved by normalizing them to the coincidences of distinguishable photons (Sec.~\ref{sec:bs}); and (iv) the four settings must probe the same input under the same conditions, i.e., the setup must be stationary during the acquisition, or the settings must be interleaved on a time scale shorter than any drift (Sec.~\ref{sec:drift}). 

Within these conditions, a simple principle guides the discussion: noise, i.e., any process that makes the two photons less similar or more mixed, cannot produce a false positive, whereas miscalibration, i.e., a violation of conditions (ii)--(iv), can bias the witness in either direction. Table~\ref{tab:imperf} summarizes the imperfections discussed below; the derivations are given in Appendices~\ref{app:robust} and \ref{app:bs}.

\begin{table}[t]
\caption{Effect of experimental imperfections on the witness. FP indicates whether a false positive is possible; the last column indicates whether the result was established analytically (A) or numerically (N).}
\label{tab:imperf}
\begin{ruledtabular}
\begin{tabular}{llll}
Imperfection & Effect & FP & \\
\hline
Imperfect indistinguishability & rescales all $\mathcal{F}_j$ & no & A\\
Nonidentical photons & witness still valid & no & A\\
Correlated photons & conditional only & no\footnotemark[1] & A\\
Common polarization drift & none & no & A\\
Relative polarization drift & decreases $\W$\footnotemark[2] & no & A\\
Setting-correlated drift & bias (up to $\frac12$) & yes & A\\
Accidental counts [$\W_V$] & common factor & no & A\\
BS asymmetry [$\W_V$] & common factor & no & A\\
BS asymmetry (naive) & bias $1-4RT$ & yes & A\\
Polarization-dependent BS & reduces $|\W|$ & no & N\\
PDL before the settings & witness still valid & no & A\\
PDL after the settings & bias $\lesssim0.19\,\delta$ & yes\footnotemark[3] & N\\
Multiphoton emission [$\W_V$] & modifies $|\W|$ & no\footnotemark[4] & A\\
Wave-plate angular errors & $|\Delta\W|\le\sum_jD_j$ & yes & A\\
Setting-dependent delay & bias $\le(1-\mathcal{F}_{\Delta t})/2$ & yes & A\\
Chromatic retardance & bias $\lesssim0.25(\sigma_\omega\tau_w)^2$ & yes\footnotemark[5] & N\\
\end{tabular}
\end{ruledtabular}
\footnotetext[1]{With respect to the conditional states; the average single-photon state is not certified (Sec.~\ref{sec:nonid}).}
\footnotetext[2]{For identical photons; for nonidentical photons the witness may increase, but cannot become positive if either photon is separable.}
\footnotetext[3]{Including polarization-dependent detection efficiencies; removed by polarization-insensitive detection (Sec.~\ref{sec:pdl}).}
\footnotetext[4]{To first order in $g^{(2)}(0)$, for extra photons independent of the main ones (Appendix~\ref{app:bs}).}
\footnotetext[5]{No false positive was found when all four settings are implemented by the same QWP--HWP--QWP sequence.}
\end{table}

\subsection{Nonidentical and correlated photons}\label{sec:nonid}

So far we have assumed that both photons are in the same state $\rho$. Fortunately, the witness remains valid even when their states differ, such as from different emission events or sources. 

For independent but nonidentical photons, in states $\rho_A$ and $\rho_B$, the two similarities become overlaps, and the witness is $\W=\tr(\rho_A\rho_B)-\tr(\rho_{A,\mathrm o}\rho_{B,\mathrm o})$, where $\rho_{A,\mathrm o}$ and $\rho_{B,\mathrm o}$ are the states of the non-polarization DOFs [Eq.~\eqref{eq:Wnonid}]. If either photon has no internal entanglement, the reduction criterion applied to its state gives $\W\le0$ (Appendix~\ref{app:robust}). A positive witness therefore certifies that \emph{both} photons are internally entangled. The converse does not hold: differences between
the photons can hide internal entanglement.
For example, if one photon is prepared in
$\ket{\Phi_+}$ and the other in $\ket{\Phi_-}$,
where
$\ket{\Phi_\pm}
=(\ket{H,t_0}\pm\ket{V,t_1})/\sqrt2$,
both photons are maximally entangled internally,
yet their orthogonality yields $\W=-1/2$.

Some care is needed when the photons are classically correlated, $\varrho_{AB}=\sum_kp_k\,\rho_{A,k}\otimes\rho_{B,k}$, as happens when a slowly fluctuating parameter of the source, such as its central frequency, takes the same value for both photons. The measured witness is then the average $\sum_kp_k\W_k$ of the values obtained for the conditional states, so a positive value shows that internally entangled conditional states occur, but \emph{not} that the average state $\bar\rho=\sum_kp_k\rho_{A,k}$ is internally entangled. For instance, if both photons are prepared in the same state $(\ket{H,t_0}\pm\ket{V,t_1})/\sqrt2$, with the sign chosen at random in each trial, then $\W=\frac12$, although the average state is separable (Appendix~\ref{app:robust}). Certifying the average state requires photons emitted farther apart in time than this correlation time.

\subsection{Polarization drift}\label{sec:drift}

Polarization drifts, e.g., due to temperature fluctuations, are common effects in optical fibers. A drift that affects both photons in the same way leaves the witness unchanged, since a common rotation changes neither the purity of the photons nor that of their other DOFs; however, relative drifts make the photons less similar and, for initially identical photons, can only \emph{decrease} the witness. This is because a random relative rotation acts as a partial depolarization of photon $B$, which increases $P_0$ but does not change the similarity measured after the polarization has been erased. For nonidentical photons it may also increase the witness, but it cannot produce a false positive, since a polarization rotation does not create internal entanglement (Sec.~\ref{sec:nonid}). Relative drift can therefore hide internal entanglement, but never fake it.

These conclusions assume that the drift is the same for the four settings. Since the settings are usually measured sequentially, a drift that is correlated with the setting acts as a setting-dependent error of the Pauli operators and can positively bias the witness (Appendix~\ref{app:robust}). The same holds for any change of the source or of the setup between settings, such as a drift of the relative delay. The remedy is to interleave or randomize the settings on a time scale shorter than any drift, so that every setting samples the same distribution of conditions; the measured witness is then an average of values to which the arguments of this section apply.

\subsection{Implementation of the Pauli operators}\label{sec:optics}

For the polarization DOF, the Pauli operators can be implemented with wave plates. In fact, a quarter-wave plate (QWP), a half-wave plate (HWP) and another QWP can implement any polarization transformation~\cite{SimonMukunda1990}, so that all four settings, including the identity, can be realized by the same optical elements, which are rotated but never removed from the beam; as discussed below, this is not merely a matter of convenience.

Errors in the orientation of the wave plates are a form of miscalibration. Since $\W$ is a combination of four probabilities, an error in one of the settings changes $\W$ by at most the trace distance between the photon states produced by the implemented and the ideal settings (Appendix~\ref{app:robust}). For a HWP with an angular error $\delta$, this is at most $\sin2\delta\approx2\delta$, so a misalignment of $0.5^\circ$ leads to $|\Delta\W|\lesssim0.02$ in the worst case. In practice the effect is usually much smaller, since errors that rotate the three axes rigidly are harmless (Sec.~\ref{sec:witness}), and, for many states, the coincidence probabilities are stationary at the ideal settings.

A more subtle requirement is that the setting optics act \emph{only} on the polarization. For example, in a scenario where a polarized photon is in a coherent superposition of two frequencies, a relative phase between the two frequency components introduced only by the optics implementing $\sigma_1$ can produce $\W>0$ (Appendix~\ref{app:robust}). 

Three effects are relevant in practice. First, a setting-dependent group delay $\Delta t$, as caused by inserting or removing wave plates, produces a bias of up to $(1-\mathcal{F}_{\Delta t})/2$, where $\mathcal{F}_{\Delta t}$ is the HOM overlap between a photon and a copy of itself delayed by $\Delta t$. For photons with picosecond coherence times, such as those generated by spontaneous parametric down-conversion, the measurement could be completely corrupted. Implementing all four settings with the same optics removes this bias. Second, the differential group delay $\tau_w$ of the wave plates couples polarization and frequency; the resulting bias is at most about $0.25(\sigma_\omega\tau_w)^2$, i.e., below $10^{-4}$ for typical quartz plates and photons with a 1-nm bandwidth, and no false positive was found when all settings use the same QWP--HWP--QWP sequence. Third, the spatial mode of photon $B$ must not depend on the setting: beam deviations caused by rotating wave plates can be removed by coupling photon $B$ into a single-mode fiber after the setting optics, the resulting polarization-independent losses being taken care of by the normalization of Sec.~\ref{sec:bs}.

\subsection{Polarization-dependent losses and detection efficiencies}\label{sec:pdl}

Polarization-dependent losses (PDL) arise in dielectric coatings and fiber components and, in particular, in superconducting nanowire detectors. A PDL acting on photon $A$, or on photon $B$ before the setting optics, transforms a photon without internal entanglement into another photon without internal entanglement, and cannot produce a false positive (Sec.~\ref{sec:nonid}). After the setting optics, however, the presence of PDL renders the ideally unpolarized photon $B$ partially polarized.

For a relative difference $\delta$ between the detection efficiencies of the most and least efficiently detected polarizations, we found false positives of up to $\simeq0.19\,\delta$ (Appendix~\ref{app:robust}); a 10\% difference thus produces a bias of 0.02, comparable to the values of $\W$ near the detection threshold (Table~\ref{tab:stats}). A possible countermeasure in this case is polarization-insensitive detection, e.g., with independent polarization scramblers in front of the two detectors, which average the efficiencies over polarizations without affecting the interference. Note that unitary polarization transformations after the setting optics are harmless: fiber birefringence between the wave plates and the BS is equivalent to the relative drift of Sec.~\ref{sec:drift}, and, for a lossless BS whose splitting ratio and reflection phase depend on the polarization, we found no false positive in our simulations, only a reduction of $|\W|$ (Appendix~\ref{app:robust}).

\subsection{Beam-splitter asymmetry and normalization}\label{sec:bs}

In a real-world BS, its reflectivity $R$ slightly differs from its transmissivity $T=1-R$. In this case, perfect bunching is never achieved, even for identical photons, and the coincidence probability becomes (Appendix~\ref{app:bs})
\begin{equation}
\Pc=R^2+T^2-2RT\,\mathcal{F}.
\end{equation}
If $\mathcal{F}$ is computed from the ideal relation $\mathcal{F}=1-2\Pc$, the extra coincidences inflate all four probabilities, and, since $\W$ adds three of them and subtracts only one, the witness obtained in this way, $\W'=4RT\,\W+(1-4RT)$, has a positive bias, e.g., $3.3\times10^{-3}$ for an asymmetry of 0.5~dB between the BS outputs, as considered in Ref.~\cite{Temporao2024}. This bias is completely removed by normalizing each measurement to the coincidence count $C_\infty$ obtained when the two photons are made fully distinguishable, e.g., by delaying one of them by much more than the coherence time. The visibilities $V_j\equiv1-C_j/C_\infty$, where $C_j$ is the coincidence count for setting $\sigma_j$, are proportional to $\mathcal{F}_j$, with a factor that does not depend on the setting, so the normalized witness
\begin{equation}
\W_V\equiv\tfrac12(V_0-V_1-V_2-V_3)
\label{eq:WV}
\end{equation}
is proportional to $\W$, with a positive constant, and has exactly the same sign; detection efficiencies and the number of pairs also cancel out. We therefore recommend $\W_V$ as the estimator to be used in experiments; if the wave plates introduce setting-dependent losses, $C_\infty$ should be measured for each setting.

\subsection{Accidental counts and multiphoton emission}

Accidental counts due to dark counts and background light are usually negligible~\cite{Temporao2024}, and are harmless for $\W_V$: since they contribute equally to $C_j$ and to $C_\infty$, they only reduce all visibilities by a common factor.

Multiphoton emission is more subtle: extra photons produce coincidences even when the main photons are perfectly identical, which mimics a reduction of their indistinguishability~\cite{Santori2002,Ollivier2021}. One might expect a positive bias, as for a naively processed BS asymmetry. This is not the case for $\W_V$, provided that $C_\infty$ is measured under the same source conditions. To first order in the probability $\varepsilon$ that a pulse contains an extra photon, with $g^{(2)}(0)\simeq2\varepsilon$, and for threshold detectors, $\W_V\le0$ for every photon without internal entanglement, whatever the state of the extra photon, in either input, provided it is independent of the main photon (Appendix~\ref{app:bs}). The reason is that the four settings together are equivalent to replacing the photon that undergoes them by a photon in the state $\Ip\otimes\rhoo-\rho$, a legitimate state for separable $\rho$, and interference with the other photons can only reduce the coincidences below those of the reference measurement. Multiphoton emission therefore only modifies the magnitude of the witness, e.g., $\W_V=\W(1-\varepsilon)/(1+\varepsilon)$ for an extra photon distinguishable from the main ones~\cite{Ollivier2021}; for an independent copy of the main photon, $\W_V$ also involves a third-order moment of $\rho$. The analysis does not cover extra photons that are correlated or entangled with the main one, contributions of order $\varepsilon^2$, or photon-number-resolving detection; in such cases, a conservative approach is to claim a violation only if $\W_V$ exceeds $g^{(2)}(0)$ by a comfortable margin.

%=====================================================================
\section{Generalizations and limitations}\label{sec:ext}
%=====================================================================

\subsection{Qudits}

The method is not restricted to polarization: the Pauli operators can act on any two-level encoding. It also extends to qudits, such as time-bin~\cite{Marcikic2003}, frequency-bin~\cite{Kues2017,Lu2019}, or orbital angular momentum~\cite{Khoury2011} encodings, which may likewise be coupled to other DOFs of the photon. The only ingredient needed is a set of operations that, applied with equal probabilities, completely scrambles the qudit while leaving the other DOFs untouched. The smallest such set has $d^2$ elements~\cite{Gross2007}, e.g., the Weyl-Heisenberg operators $X^aZ^b$, with $a,b=0,\dots,d-1$, built from the cyclic shift $X$, which moves the photon from level $k$ to level $k+1$ modulo $d$, and the phase ramp $Z$, which multiplies level $k$ by $e^{2\pi ik/d}$. The witness is then (Appendix~\ref{app:qudit})
\begin{equation}
\W_d=\frac2d\sum_{a,b=0}^{d-1}P_{ab}-2P_{00}-(d-1)=\tr\rho^2-\tr\rhoo^2,
\label{eq:Wd}
\end{equation}
where $P_{ab}$ is the coincidence probability for the setting $X^aZ^b$; for $d=2$, Eq.~\eqref{eq:Wd} reduces to Eq.~\eqref{eq:W}. For identical photons, $-(1-1/d)\le\W_d\le1-1/d$, the upper bound being reached when the qudit is maximally entangled with the other DOFs. Experimental implementations for this case are outside the scope of the paper. With the Weyl-Heisenberg operators, the number of settings grows only as $d^2$, whereas a tomographic test would require the full joint state, including the resolution of the other DOFs.

\subsection{Limitations}\label{sec:limits}

Finally, let us discuss the limitations of the method.
\\

\emph{The witness is a sufficient, but not necessary, condition for mixed states.} As illustrated by the orange region of Fig.~\ref{fig:regions}, some internally entangled photons are not detected, since $\W$ is built from only two numbers, the purity of the whole and that of the non-polarization part, whereas entanglement is a property of the full joint state. The complementary criterion $\W_{\mathrm p}$ detects some states that $\W$ misses, and vice versa (Sec.~\ref{sec:qd}), and purity-based lower bounds on entanglement measures~\cite{MintertBuchleitner2007,Walborn2006} could also be evaluated; even then, entanglement can be hidden from both criteria, as for the states of Sec.~\ref{sec:mixed}, for which $\W_{\mathrm p}$ coincides with $\W$. The ultimate limit is full tomography of the joint state, which, however, requires resolving all DOFs.

\emph{The coupled DOF is not identified.} By construction, $\Ho$ collects \emph{all} internal DOFs other than polarization, so a positive witness says that polarization is entangled with \emph{some} of them. Additional information, such as from a delay-resolved measurement, or its frequency counterpart, could be used to identify the DOF with which the polarization is correlated, and the witness to certify that this correlation involves entanglement.

\emph{The DOP is not available in the general mixed case.} The HOM coincidences measure the mode-resolved Stokes operators, $\norm{A_j}^2$, rather than the mode-averaged Stokes parameters, $\tr A_j$ (Sec.~\ref{sec:method}), so the formulas of Ref.~\cite{Temporao2024} for the DOP hold only for pure photons or for photons whose other DOFs are in a pure state uncorrelated with the polarization. The witness $\W$ does not require the DOP, but the complementary criterion $\W_{\mathrm p}$ does, and must then rely on conventional polarimetry.

\emph{The two photons must be independent.} All the results of this paper assume that the two photons are independent or, at most, classically correlated, as for photons from different emission events or different sources; in the latter case, the certificate refers to their conditional states (Sec.~\ref{sec:nonid}). If the photons are entangled \emph{with each other}, as for the two photons of a pair generated by spontaneous parametric down-conversion, the HOM effect probes the exchange symmetry of their joint state, which depends on the correlations between them and not only on their individual states. The witness can then take any value between $-2$ and $2$ (Appendix~\ref{app:proof}): the polarization singlet with identical other DOFs, which always produces coincidences~\cite{HorodeckiRMP2009}, gives $\W=-2$, whereas the product of a polarization singlet and a time-bin singlet gives $\W=2$, although each photon alone is in a maximally mixed, and hence separable, state. Since independent or classically correlated photons always give $-1\le\W\le\frac12$ (Sec.~\ref{sec:witness}), a value outside this interval reveals, within the conditions of Sec.~\ref{sec:imperf}, that the photons are entangled with each other; inside it, however, no such guarantee exists, and independence must be ensured by the design of the experiment. This deserves attention for emitters with a long-lived memory: in a charged QD, for instance, successive photons can become entangled with each other through the spin of the resident carrier, which is the basis of photonic cluster-state generation~\cite{LindnerRudolph2009}.

%=====================================================================
\section{Conclusion}\label{sec:concl}
%=====================================================================

We have proposed a new experimental scheme, based on two-photon interference, for measuring a witness that can detect entanglement between the polarization of a single photon and its other internal DOFs, even when the photon is also entangled with an external environment or subject to random polarization rotations common to both photons. Averaging four HOM measurements, with the identity and the three Pauli operators applied to one of the photons, is equivalent to completely depolarizing that photon, which yields the purity of the non-polarization DOFs without resolving them. Comparing it with the purity of the photon as a whole implements a known separability criterion in the form of the nonlinear witness $\W=P_1+P_2+P_3-P_0-1$, which requires no polarization-selective components, no knowledge of the Stokes parameters or even of the DOP, and no alignment of the measurement axes; being a sufficient condition, it can miss weakly entangled mixed states.

For quantum-dot photons, the witness certifies the polarization--frequency entanglement induced by the fine-structure splitting in the presence of dephasing, and rejects the classically correlated photons of the biexciton cascade. Unlike polarimetry resolved in time or in frequency, whose statistics are always compatible with a separable photon, it provides this certificate without temporal or spectral resolution and without a model of the source. Combined with a conventional measurement of the DOP, the global purity also yields a complementary criterion, which is more sensitive for nearly pure photons.

For independent photons, a positive witness certifies the internal entanglement of the individual photon states. For classically correlated photons, it certifies internal entanglement in conditional components of the preparation, but not necessarily in the averaged single-photon state. Photons entangled with each other are not covered, since they can produce positive values of the witness without any internal entanglement. In either case, reliable certification requires setting optics acting exclusively on polarization, no polarization-dependent loss after them, including at detection, statistically equivalent acquisition conditions for the four settings, and proper coincidence normalization. Within these conditions, we believe that the method can be useful for characterizing single-photon sources in which several depolarization mechanisms coexist, such as quantum dots, and for studying polarization-mode dispersion in quantum communication links.

\section*{Data availability}
The code and data used to generate all figures and numbers of this paper, including the quantum circuits, their independent density-matrix benchmark, the models of Appendices~\ref{app:robust}, \ref{app:bs}, and \ref{app:qdmodel}, and a script that recomputes and checks every number quoted in the text, will be deposited in \href{https://www.maxwell.vrac.puc-rio.br}{PUC-Rio's Maxwell repository} upon publication; until then, they are available from the author upon reasonable request.

\section*{AI Use Disclosure}

The author used GPT-6 (OpenAI, through ChatGPT) and
Claude Opus 5.5 (Anthropic) to assist with mathematical
reasoning, numerical code development, figure generation,
and manuscript preparation. The author reviewed and
corrected the AI-assisted outputs and checked the
reported analytical and numerical results against
independent calculations and density-operator methods.
The author assumes full responsibility for the scientific
content and accuracy of the manuscript.

\begin{acknowledgments}
The author acknowledges financial support from Conselho Nacional de Desenvolvimento Científico e Tecnológico (CNPq), grant number 305034/2026-0. The author thanks F. P. Guids for the helpful discussions.
\end{acknowledgments}

\appendix

%=====================================================================
\section{Coincidence probability for mixed and correlated inputs}\label{app:hom}
%=====================================================================

For two photons with pure internal states $\ket a$ and $\ket b$ impinging on a balanced BS, the coincidence probability is $\frac12(1-|\braket ab|^2)$ (Appendix~B of Ref.~\cite{Temporao2024}). This can be written as $\bra a\bra b\Pi_-\ket a\ket b$, where $S$ is the swap operator, $S\ket a\ket b=\ket b\ket a$, which exchanges the internal states of the two photons, and $\Pi_-=\frac12(\mathbb I-S)$ is the projector onto the antisymmetric subspace; this is the equivalence between the HOM effect and the swap test~\cite{GarciaEscartin2013}. Since the coincidence probability is linear in the two-photon input, for an arbitrary (possibly correlated) two-photon internal state $\varrho_{AB}$ one has
\begin{equation}
\Pc=\tr(\varrho_{AB}\Pi_-).
\label{eq:swap}
\end{equation}
For independent photons, $\varrho_{AB}=\rho_A\otimes\rho_B$, and the identity $\tr[(X\otimes Y)S]=\tr(XY)$ gives $\Pc=\frac12[1-\tr(\rho_A\rho_B)]$. Setting $\rho_A=\rho$ and $\rho_B=(U\otimes\Io)\rho(U\otimes\Io)^\dagger$ yields Eq.~\eqref{eq:F}. For pure states, Eq.~\eqref{eq:F} reduces to Eqs.~(2) and (3) of Ref.~\cite{Temporao2024}.

%=====================================================================
\section{Stokes operators and the Pauli sum rule}\label{app:stokes}
%=====================================================================

Expanding the polarization part of $\rho$ in the Pauli basis gives
\begin{equation}
\rho=\frac12\sum_{i=0}^{3}\sigma_i\otimes A_i,
\label{eq:decomp}
\end{equation}
with $A_i$ defined in Eq.~\eqref{eq:A}. Writing $\st i\equiv\sigma_i\otimes\Io$ and using $\sigma_j\sigma_i\sigma_j=\varepsilon_{ji}\sigma_i$, where $\varepsilon_{ji}=+1$ if $j=0$, $i=0$, or $j=i$, and $\varepsilon_{ji}=-1$ otherwise, one obtains $\st j\rho\,\st j=\frac12\sum_i\varepsilon_{ji}\sigma_i\otimes A_i$. Since $\tr[(\sigma_\kappa\otimes A_\kappa)(\sigma_i\otimes A_i)]=2\delta_{\kappa i}\tr A_i^2$,
\begin{equation}
\mathcal{F}_j=\frac12\sum_{i}\varepsilon_{ji}\norm{A_i}^2,
\label{eq:Fa}
\end{equation}
which shows that the coincidence probabilities depend only on the norms of the Stokes operators. The matrix $\varepsilon$ is symmetric and satisfies $\varepsilon^2=4\mathbb I_4$, so Eq.~\eqref{eq:Fa} can be inverted, $\norm{A_i}^2=\frac12\sum_j\varepsilon_{ij}\mathcal{F}_j$, and, using $\mathcal{F}_j=1-2P_j$ [Eq.~\eqref{eq:F}], expressed in terms of the coincidence probabilities:
\begin{equation}
\begin{aligned}
\norm{A_0}^2&=2-(P_0+P_1+P_2+P_3),\\
\norm{A_1}^2&=P_2+P_3-P_0-P_1,\\
\norm{A_2}^2&=P_1+P_3-P_0-P_2,\\
\norm{A_3}^2&=P_1+P_2-P_0-P_3.
\end{aligned}
\label{eq:aP}
\end{equation}
Since the left-hand sides are nonnegative, physical data must satisfy three linear constraints, which provide a useful consistency check against systematic errors.

For a pure state, Eq.~\eqref{eq:Fa} gives $\mathcal{F}_j=\langle s_j\rangle^2$, and, for $\rho=\rhop\otimes\ket\xi\!\bra\xi$, $A_j=\langle s_j\rangle\ket\xi\!\bra\xi$ and $\mathcal{F}_j=\langle s_j\rangle^2+\frac12(1-\mathrm{DOP}^2)$, which are Eqs.~(6) and (14) of Ref.~\cite{Temporao2024}; for product states $\rho=\rhop\otimes\rhoo$, all fidelities are multiplied by $\tr\rhoo^2$.

For any operator $X$ acting on $\Hp\otimes\Ho$, the Pauli expansion of Eq.~\eqref{eq:decomp} and $\sum_i\varepsilon_{ij}=4\delta_{j0}$ give the Pauli twirl, a standard unitary-design identity~\cite{Gross2007},
\begin{equation}
\frac14\sum_{i=0}^3\st i X\st i=\frac{\Ip}{2}\otimes\tr_{\mathrm p}X ,
\label{eq:twirl}
\end{equation}
of which Eq.~\eqref{eq:twirlrho} is the case $X=\rho$. Equivalently, for the swap $S=S_{\mathrm p}\otimes S_{\mathrm o}$ of the two photons, with $S_{\mathrm p}=\frac12\sum_j\sigma_j\otimes\sigma_j$, one has $\sum_{i=0}^3(\mathbb I\otimes\sigma_i)S_{\mathrm p}(\mathbb I\otimes\sigma_i)=2\,\mathbb I_{\mathrm{pp}}$, where $\mathbb I_{\mathrm{pp}}$ is the identity acting on the polarizations of the two photons: the average over the four settings is a swap test on the non-polarization DOFs.

%=====================================================================
\section{The witness and its bounds}\label{app:proof}
%=====================================================================

\emph{Separability criteria.} For a separable state, $\rho=\sum_kp_k\,\alpha_k\otimes\beta_k$, the reduction criterion~\cite{HorodeckiReduction1999,Cerf1999} states that
\begin{equation}
\Delta\equiv\Ip\otimes\rhoo-\rho\ge0,
\label{eq:Delta}
\end{equation}
and, similarly, $\rhop\otimes\Io-\rho\ge0$. Since $\tr(XY)\ge0$ for positive semidefinite operators $X$ and $Y$, the expectation value of $\Delta$ in $\rho$ itself is nonnegative for separable states,
\begin{equation}
\tr(\rho\Delta)=\tr\rhoo^2-\tr\rho^2=-\W\ge0,
\end{equation}
which is Eq.~\eqref{eq:ineq}; the same inequality also follows from the entropic and majorization criteria of Refs.~\cite{HorodeckiEntropy1996,NielsenKempe2001}. In the same way, $\rhop\otimes\Io-\rho\ge0$ gives $\tr\rho^2\le\tr\rhop^2$, and hence the criterion $\W_{\mathrm p}$ of Eq.~\eqref{eq:Wp}. The operator inequality~\eqref{eq:Delta} is a stronger test than Eq.~\eqref{eq:ineq}: since polarization is a qubit, it is equivalent to the partial-transpose criterion~\cite{HorodeckiReduction1999,HorodeckiRMP2009}, which detects every entangled state when $\dim\Ho\le3$. The witness, however, uses only one number, the expectation value $\tr(\rho\Delta)$, which can be positive even when $\Delta$ has a negative eigenvalue, if $\rho$ has little weight on the corresponding eigenvector. For the states of Sec.~\ref{sec:mixed} with $D_0=0$, for instance, $\rhoo=\Io/2$, and $\Delta$ has the eigenvalue $(1-3q)/4$ along $\ket\Psi$, which is negative for $q>\frac13$, whereas $\tr(\rho\Delta)=(1-3q^2)/4$ becomes negative only for $q>1/\sqrt3$; this is the origin of the orange region in Fig.~\ref{fig:regions}.

\emph{Bounds for identical photons.} Applying Eq.~\eqref{eq:twirl} to $X=\rho$ gives $2\,\Ip\otimes\rhoo-\rho=\sum_{i=1}^3\st i\rho\,\st i\ge0$, i.e., $\rho\le2\,\Ip\otimes\rhoo$ for every state, the analog of Eq.~\eqref{eq:Delta} with the factor 1 replaced by 2, the dimension of $\Hp$. Taking the expectation value in $\rho$ gives $\tr\rho^2\le2\tr\rhoo^2$, and, since $\tr\rho^2\le1$,
\begin{equation}
\W\le\min\bigl(\tr\rhoo^2,\,1-\tr\rhoo^2\bigr)\le\tfrac12,
\end{equation}
with equality for maximally entangled pure states, such as $(\ket{H,t_0}+\ket{V,t_1})/\sqrt2$. Since Eq.~\eqref{eq:decomp} gives $\tr\rho^2=\frac12\sum_{j=0}^3\norm{A_j}^2$ and $\tr\rhoo^2=\norm{A_0}^2$, the witness is $\W=\frac12(\norm{A_1}^2+\norm{A_2}^2+\norm{A_3}^2-\norm{A_0}^2)$: the length of the mode-resolved Stokes vector, $(\norm{A_1}^2+\norm{A_2}^2+\norm{A_3}^2)^{1/2}$, never exceeds $\norm{A_0}$ for separable states, in analogy with the condition $\mathrm{DOP}\le1$, but can reach $\sqrt3\,\norm{A_0}$ for internally entangled ones. For the lower bound, the blocks $\rho_{xy}=\bra x\rho\ket y$, with $x,y\in\{H,V\}$, satisfy $\rhoo=\rho_{HH}+\rho_{VV}$ and $\tr\rho^2=\sum_{x,y}\norm{\rho_{xy}}^2$, where $\norm X^2\equiv\tr(X^\dagger X)$; by the parallelogram law, $\tr\rhoo^2\le2(\norm{\rho_{HH}}^2+\norm{\rho_{VV}}^2)\le2\tr\rho^2$, so that
\begin{equation}
\W\ge-\tfrac12\tr\rhoo^2\ge-\tfrac12,
\end{equation}
with equality for $(\Ip/2)\otimes\ket\xi\!\bra\xi$, i.e., for purely external depolarization. The same steps, with the Weyl-Heisenberg twirl, give the qudit bounds of Appendix~\ref{app:qudit}.

\emph{Independent, nonidentical photons.} For photons in states $\rho_A$ and $\rho_B$, with the settings applied to photon $B$, $\mathcal F_j=\tr(\rho_A\st j\rho_B\st j)$. The identity setting gives the overlap between the two photons, $\mathcal F_0=\tr(\rho_A\rho_B)$, and Eq.~\eqref{eq:twirl}, applied to $\rho_B$, gives $\frac12\sum_j\mathcal F_j=\tr(\rho_{A,\mathrm o}\rho_{B,\mathrm o})$, the overlap between their non-polarization parts, with $\rho_{X,\mathrm o}=\tr_{\mathrm p}\rho_X$. Hence
\begin{equation}
\W=\tr(\rho_A\rho_B)-\tr(\rho_{A,\mathrm o}\rho_{B,\mathrm o}),
\label{eq:Wnonid}
\end{equation}
which reduces to $\tr\rho^2-\tr\rhoo^2$ for identical photons. Both overlaps lie between 0 and 1, and $\rho_B\le2\,\Ip\otimes\rho_{B,\mathrm o}$ gives $\tr(\rho_A\rho_B)\le2\tr(\rho_{A,\mathrm o}\rho_{B,\mathrm o})$, so that, as for identical photons, $\W\le\frac12$, whereas $\tr(\rho_A\rho_B)\ge0$ gives $\W\ge-1$. The lower bound is reached by $\rho_A=\ket{H,\xi}\!\bra{H,\xi}$ and $\rho_B=\ket{V,\xi}\!\bra{V,\xi}$, which are orthogonal in polarization and identical in all other DOFs. For classically correlated photons, $\varrho_{AB}=\sum_kp_k\,\rho_{A,k}\otimes\rho_{B,k}$, the coincidence probabilities are linear in $\varrho_{AB}$ [Eq.~\eqref{eq:swap}], so $\W$ is the average of the values obtained for the pairs $(\rho_{A,k},\rho_{B,k})$, and $-1\le\W\le\frac12$ holds for every two-photon state that is separable with respect to the two photons.

\emph{Arbitrary two-photon states.} If the photons are entangled with each other, $P_j=\tr(\varrho_{AB}^{(j)}\Pi_-)$ [Eq.~\eqref{eq:swap}], where $\varrho_{AB}^{(j)}$ is the state after setting $\sigma_j$ is applied to photon $B$. With $S=S_{\mathrm p}\otimes S_{\mathrm o}$ and $\sum_j(\mathbb I\otimes\sigma_j)S_{\mathrm p}(\mathbb I\otimes\sigma_j)=2\,\mathbb I_{\mathrm{pp}}$ (Appendix~\ref{app:stokes}), one finds $\sum_jP_j=2-\langle S_{\mathrm o}\rangle$, where $\langle S_{\mathrm o}\rangle=\tr[\varrho_{AB}(\mathbb I_{\mathrm{pp}}\otimes S_{\mathrm o})]$ is the exchange symmetry of the other DOFs. Since $P_0=\frac12(1-\langle S\rangle)$,
\begin{equation}
\W=\langle S\rangle-\langle S_{\mathrm o}\rangle=-2\tr[\varrho_{AB}(\Pi^-_{\mathrm p}\otimes S_{\mathrm o})],
\label{eq:Wswap}
\end{equation}
where $\Pi^-_{\mathrm p}=\frac12(\mathbb I_{\mathrm{pp}}-S_{\mathrm p})$ projects onto the polarization singlet: the witness compares the exchange symmetry of the photons as a whole with that of their non-polarization parts, and reduces to Eq.~\eqref{eq:Wnonid} for independent photons. Since $\Pi^-_{\mathrm p}\otimes S_{\mathrm o}$ has eigenvalues $0$ and $\pm1$, $-2\le\W\le2$, and both extremes require a polarization singlet. The singlet with identical other DOFs gives $P_0=1$ and $P_1=P_2=P_3=0$, i.e., $\W=-2$; the product of a polarization singlet and a time-bin singlet gives $P_0=0$ and $P_1=P_2=P_3=1$, i.e., $\W=2$, although each photon, taken alone, is then in the separable state $\mathbb I_4/4$. A value outside $[-1,\frac12]$ therefore reveals entanglement between the photons, in which case $\W$ no longer certifies internal entanglement.

%=====================================================================
\section{Robustness}\label{app:robust}
%=====================================================================

\emph{Nonidentical photons.} For independent photons in states $\rho_A$ and $\rho_B$, with the settings applied to photon $B$, the witness is given by Eq.~\eqref{eq:Wnonid}, $\W=\tr(\rho_A\rho_B)-\tr(\rho_{A,\mathrm o}\rho_{B,\mathrm o})$. If $\rho_B$ is separable, the reduction criterion gives $\tr[\rho_A(\Ip\otimes\rho_{B,\mathrm o}-\rho_B)]\ge0$, i.e., $\W\le0$; if $\rho_A$ is separable, the roles of the photons are exchanged. For classically correlated photons, $\varrho_{AB}=\sum_kp_k\rho_A^{(k)}\otimes\rho_B^{(k)}$, Eq.~\eqref{eq:swap} shows that $\W$ is the average of the corresponding individual values, each of which is nonpositive if one member of each pair is separable. This does not extend to the average state $\bar\rho=\sum_kp_k\rho_A^{(k)}$: for $\varrho_{AB}=\frac12(\rho_+\otimes\rho_++\rho_-\otimes\rho_-)$, with $\rho_\pm=\ket{\Phi_\pm}\!\bra{\Phi_\pm}$ and $\ket{\Phi_\pm}=(\ket{H,t_0}\pm\ket{V,t_1})/\sqrt2$, each of the three Pauli operators maps $\ket{\Phi_\pm}$ onto an orthogonal state, so $P_0=0$, $P_1=P_2=P_3=\frac12$, and $\W=\frac12$; the average state $\bar\rho=\frac12(\ket{H,t_0}\!\bra{H,t_0}+\ket{V,t_1}\!\bra{V,t_1})$, on the other hand, is separable, with $\tr\bar\rho^2=\tr\bar\rho_{\mathrm o}^2=\frac12$, i.e., $\W(\bar\rho)=0$.

\emph{Relative polarization drift.} A polarization rotation that acts equally on both photons, and does not depend on the setting, leaves $\W$ unchanged: it only rotates the three Pauli axes, which changes $\mathcal{F}_1$, $\mathcal{F}_2$, and $\mathcal{F}_3$ individually, but neither $\mathcal{F}_0$ nor $\sum_j\mathcal{F}_j$. A random rotation $V$ of photon $B$ alone, with the same distribution for all four settings, leaves $\rhoo$, and hence $\sum_j\mathcal{F}_j=2\tr\rhoo^2$, unchanged, whereas, by the Cauchy-Schwarz inequality, $\mathcal{F}'_0=\mathbb E_V\tr(\rho\tilde V\rho\tilde V^\dagger)\le\tr\rho^2$, with $\tilde V=V\otimes\Io$. For initially identical photons, a relative drift can therefore only reduce the witness, $\W'\le\W$, in agreement with the principle stated in Sec.~\ref{sec:imperf}. For nonidentical photons it may increase the witness, e.g., from $-1$ to 0 for $\rho_A=\ket{H,t_0}\!\bra{H,t_0}$, $\rho_B=\ket{V,t_0}\!\bra{V,t_0}$, and $V=\sigma_2$, but it cannot produce a false positive, since a polarization rotation, or a mixture of rotations, preserves separability, and the result of the previous paragraph applies.

\emph{Setting-correlated drift.} The results above assume that each setting samples the same distribution of states, either because the setup is stationary or because the drift is independent of the setting. When the settings are measured in consecutive blocks, a drift that occurs during the acquisition, for instance a slow temperature change in a fiber, can instead be correlated with the setting, so that different settings probe different states. This can produce false positives even for separable photons. As a first example, consider two photons in the separable state $\ket{H,t_0}$, and suppose that the polarization of photon $B$ is rotated by $V_1=\exp(-i\epsilon\sigma_3/2)$, i.e., that its linear polarization is rotated by an angle $\epsilon/2$, but only while $\sigma_1$ is measured. Then $\mathcal{F}_0=1$, $\mathcal{F}_1=|\bra H\sigma_1V_1\ket H|^2=\cos^2(\epsilon/2)$, and $\mathcal{F}_2=\mathcal{F}_3=0$, so that
\begin{equation}
\W=\tfrac12(\mathcal{F}_0-\mathcal{F}_1-\mathcal{F}_2-\mathcal{F}_3)=\tfrac12\sin^2(\epsilon/2)>0 .
\end{equation}
Had the same rotation acted during all four settings, the photons would simply be nonidentical, with $\mathcal{F}_0=\mathcal{F}_1=\cos^2(\epsilon/2)$ and $\mathcal{F}_2=\mathcal{F}_3=\sin^2(\epsilon/2)$, so that $\W=-\sin^2(\epsilon/2)\le0$, in agreement with the previous paragraphs: the false positive arises only because the rotation is correlated with the setting. The effect is small, e.g., $\W\approx0.005$ for $\epsilon=0.2$, i.e., for a rotation of about $6^\circ$, but, being systematic, it does not decrease with the number of detected pairs. A change of the source itself can have a much larger effect. If the source emits photons in the state $\ket{{+}45,t_0}$ while $\sigma_1$ is measured, and in $\ket{H,t_0}$ while $\sigma_2$ and $\sigma_3$ are measured, then $\mathcal{F}_0=1$, whichever state is emitted while $\sigma_0$ is measured, and $\mathcal{F}_1=\mathcal{F}_2=\mathcal{F}_3=0$, so that $\W=\frac12$: the largest value attainable by identical photons is obtained although every state is separable, because the four fidelities refer to different states.

Both effects are avoided by interleaving or randomizing the settings, so that the choice of setting is independent of the drift. Let $\lambda$ label the drift realization, i.e., the state of the source and of the optical paths at a given time. If the setting is chosen independently of $\lambda$, every setting samples the same distribution of $\lambda$, the measured fidelities are averages, $\mathcal{F}_j=\mathbb E_\lambda\mathcal{F}_j^{(\lambda)}$, and, since $\W$ is linear in the fidelities,
\begin{equation}
\W=\mathbb E_\lambda\W^{(\lambda)},
\end{equation}
where $\W^{(\lambda)}$ is the value that the witness would take if the drift were frozen at $\lambda$ during the whole measurement. For each $\lambda$, the four settings probe the same pair of photon states, so $\W^{(\lambda)}$ obeys the results of the previous paragraphs; in particular, $\W^{(\lambda)}>0$ requires both photons to be internally entangled for that realization. A positive average therefore still certifies internal entanglement, which is the content of condition (iv) of Sec.~\ref{sec:imperf}.

\emph{Wave-plate errors.} A misaligned wave plate implements $U_j=\sigma_jR_j$ instead of $\sigma_j$, i.e., the ideal setting preceded by a small polarization rotation $R_j$ of photon $B$, acting in that setting only. Since the probability of any event changes by at most the trace distance between the states on which it is measured, $|\Delta P_j|\le D_j$, where $D_j$ is the trace distance between $\rho$ and $\tilde R_j\rho\tilde R_j^\dagger$, with $\tilde R_j=R_j\otimes\Io$. Since the identity setting involves no optics, $P_0$ is unaffected, and
\begin{equation}
|\Delta\W|\le D_1+D_2+D_3 .
\end{equation}
For a rotation by an angle $\alpha$ of the Poincar\'e sphere, $D_j\le\sin(\alpha/2)$, since $|\bra\Psi\tilde R_j\ket\Psi|\ge\cos(\alpha/2)$ for every pure state and the trace distance is convex. A HWP intended to implement $\sigma_1$ but rotated by $\delta$ implements $U_1=\cos2\delta\,\sigma_1+\sin2\delta\,\sigma_2=\sigma_1R_1$, where $R_1$ is a rotation by $\alpha=4\delta$ about the $\sigma_3$ axis, so that $D_1\le\sin2\delta\approx2\delta$, and a misalignment of $0.5^\circ$ gives $|\Delta\W|\lesssim0.02$. For $\sigma_3$, which requires two plates, the errors of the two plates add. Since the argument involves only the photon states produced by the actual and ideal settings, it applies to any imperfection of the setting optics, including effects on the other DOFs.

\emph{Setting optics acting on other DOFs.} For the example of Sec.~\ref{sec:optics}, $\rho=\ket H\!\bra H\otimes\ket\xi\!\bra\xi$ with $\ket\xi=(\ket{\omega_1}+\ket{\omega_2})/\sqrt2$, and the $\sigma_1$ setting implemented as $\sigma_1\otimes\mathrm{diag}(1,e^{i\phi})$. Then $\mathcal{F}_0=1$, $\mathcal{F}_1=|\langle\xi|\mathrm{diag}(1,e^{i\phi})|\xi\rangle|^2=\cos^2(\phi/2)$, and $\mathcal{F}_2=\mathcal{F}_3=0$, so that $\W=\frac12(\mathcal{F}_0-\mathcal{F}_1)=\frac12\sin^2(\phi/2)$. More generally, if setting $\sigma_j$ also delays photon $B$ by $\Delta t_j$ with respect to the identity setting, a product state $\ket{\bm n}\!\bra{\bm n}\otimes\rhoo$ with pure polarization along the unit Stokes vector $\bm n$ gives $\mathcal{F}_0=\tr\rhoo^2$ and $\mathcal{F}_j=n_j^2\mathcal{F}_{\Delta t_j}$, where $\mathcal{F}_{\Delta t}$ is the overlap between $\rhoo$ and its delayed copy; hence $\W=\frac12\sum_jn_j^2(\tr\rhoo^2-\mathcal{F}_{\Delta t_j})$, which, for pure $\rhoo$, is bounded by $\frac12\max_j(1-\mathcal{F}_{\Delta t_j})$. For a Gaussian spectrum with rms width $\sigma_\omega$, $\mathcal{F}_{\Delta t}=\exp(-\sigma_\omega^2\Delta t^2)$; for the exponential wave packets of Appendix~\ref{app:qdmodel}, $\mathcal{F}_{\Delta t}=\exp(-\Gamma|\Delta t|)$.

\emph{Chromatic retardance.} We modeled each wave plate by a retarder whose retardance depends linearly on the frequency, $\Gamma_{\mathrm H}(\omega)=\pi+\tau_w(\omega-\omega_0)$ for the HWP and $\Gamma_{\mathrm Q}(\omega)=\pi/2+\tau_w(\omega-\omega_0)/2$ for the QWPs, where $\tau_w$ is the differential group delay of the HWP; the common group delay was assumed to be compensated. For a photon with a Gaussian spectrum of rms width $\sigma_\omega$ and pure polarization, uncorrelated with the frequency, we computed $\W$ numerically for polarizations on a grid covering the Poincar\'e sphere. When the identity setting is implemented without optics and the Pauli settings by single HWPs (or by a QWP--HWP--QWP sequence), the largest value of $\W$ is about $0.25(\sigma_\omega\tau_w)^2$ for $\sigma_\omega\tau_w\le0.3$. When all four settings, including the identity, are implemented by the same QWP--HWP--QWP sequence, $\W$ was negative for all polarizations, between $-0.38(\sigma_\omega\tau_w)^2$ and $-0.24(\sigma_\omega\tau_w)^2$.

\emph{Polarization-dependent losses.} A PDL is described by a filter $K=k\otimes\Io$, with $0\le k^\dagger k\le\Ip$. If it acts on photon $A$, or on photon $B$ before the setting optics, it maps a separable state onto (the normalized version of) another separable state, $K\rho K^\dagger/\tr(K\rho K^\dagger)$, and the result for nonidentical photons applies. A polarization-dependent detection efficiency is described by an efficiency operator $E=e\otimes\Io$, the same for both detectors. For pure states $\ket a$ and $\ket b$ the coincidence probability becomes $\frac12[\bra aE\ket a\bra bE\ket b-|\bra aE\ket b|^2]$, so the normalized visibility measures the overlap between the filtered states $E^{1/2}\ket a$ and $E^{1/2}\ket b$; since $E^{1/2}$ acts on photon $B$ after the setting, this amounts to a setting-dependent filter, $\sigma_jE^{1/2}\sigma_j$, acting on photon $B$ before the setting. We computed $\W_V$, with $C_\infty$ measured for each setting, for $e=\mathrm{diag}(1,1-\delta)$, and maximized it over separable polarization--time-bin states (mixtures of two or three product states) by repeated local optimization. The maximum was $1.9\times10^{-3}$, $5.8\times10^{-3}$, and $1.9\times10^{-2}$ for $\delta=0.01$, $0.03$, and $0.1$, respectively, i.e., $\simeq0.19\,\delta$. A PDL with $\delta=0.1$ acting on photon $B$ between the setting optics and the BS gave $2.0\times10^{-2}$, whereas no positive value was found for a PDL acting on photon $A$ or on photon $B$ before the setting optics. Independent polarization scramblers in front of the two detectors replace $e$ by its polarization average, $\frac12(\tr e)\,\Ip$, which restores the ideal behavior up to a common factor. Finally, for a lossless BS whose reflectivity and reflection phase depend on the polarization ($R_{H,V}=\frac12\pm0.05$ and a relative reflection phase of 0.3~rad), the same optimization gave a maximum of $2\times10^{-16}$, i.e., no false positive, while $\W_V$ for a maximally entangled state decreased from 0.50 to 0.47. This is a numerical observation for the configurations studied, not a general theorem.

%=====================================================================
\section{Beam-splitter asymmetry and multiphoton emission}\label{app:bs}
%=====================================================================

For a lossless beam splitter with intensity transmittance $T$ and reflectance $R$, the standard Hong--Ou--Mandel coincidence probability is
\begin{equation}
\Pc=R^2+T^2-2RT\,\mathcal{F},
\end{equation}
where $\mathcal{F}$ is the internal-state overlap~\cite{HOM1987,Uppu2016}. With $R+T=1$, the naive estimate $\mathcal{F}'=1-2\Pc=4RT(1+\mathcal{F})-1$ then leads to $\W'=\frac12(\mathcal{F}'_0-\sum_{j=1}^3\mathcal{F}'_j)=4RT\W+(1-4RT)$. With distinguishable photons, $\mathcal{F}=0$ and $C_\infty\propto R^2+T^2$, so that $V_j=1-C_j/C_\infty=\kappa\mathcal{F}_j$, with $\kappa=2RT/(R^2+T^2)$, and $\W_V=\kappa\W$. For a lossy BS, or with a constant accidental background $c$, which gives $V_j=(C_\infty-C_j)/(C_\infty+c)$, the proportionality factor changes, but it is the same for all settings, so that $\W_V$ remains proportional to $\W$.

\emph{Multiphoton emission.} Let each input pulse contain, with probability $\varepsilon$, an extra photon, whose internal state, $\nu_A$ in input $A$ or $\nu_B$ in input $B$, is independent of that of the main photon, and consider a balanced BS and threshold detectors, so that $g^{(2)}(0)\simeq2\varepsilon$. To first order in $\varepsilon$, three kinds of events contribute to the coincidences: two-photon events, with probability $1-2\varepsilon$, and three-photon events, in which the extra photon is in input $A$ or in input $B$, each with probability $\varepsilon$. An extra photon in input $B$ is also subjected to the setting.

For two photons in different inputs, with internal states $\tau_A$ and $\tau_B$, Eq.~\eqref{eq:swap} gives $\Pc=\frac12[1-\tr(\tau_A\tau_B)]$. Three-photon events obey a formula of the same form. Let photons 1 and 2, with internal states $\tau_1$ and $\tau_2$, enter the BS through the same input, and photon 3, with internal state $\tau_3$, through the other. The two photons sharing an input mode occupy the symmetric subspace: their joint internal state is $\omega_{12}\propto\Pi_{12}(\tau_1\otimes\tau_2)\Pi_{12}$, where $\Pi_{12}=\frac12(I+S_{12})$ and $S_{kl}$ swaps the internal states of photons $k$ and $l$. The reduced state of either photon, obtained by symmetrizing $\tau_1\otimes\tau_2$ and tracing out one photon, is
\begin{equation}
\omega=\frac{\tau_1+\tau_2+\tau_1\tau_2+\tau_2\tau_1}{2[1+\tr(\tau_1\tau_2)]},
\end{equation}
following the standard symmetric-projection construction~\cite{Barenco1997}.
The three-photon coincidence probability for a balanced BS with input occupation $(2,1)$ has been derived for Gaussian wave packets that differ by a delay~\cite{PradanaChew2019}. For arbitrary mixed internal states, it follows from the fact that the three photons can leave through the same output only in their symmetric subspace, so that the probability of this event is proportional to $\tr[\Pi_{123}(\omega_{12}\otimes\tau_3)]$, where $\Pi_{123}$ is the symmetric projector of the three photons. On states that are symmetric in photons 1 and 2, $\Pi_{123}=\frac13(I+S_{13}+S_{23})$, and each swap gives $\tr(\omega\tau_3)$, by the identity that leads to Eq.~\eqref{eq:swap}. Fixing the constant with a distinguishable photon 3, for which the three photons leave through the same output with probability $\frac14$, the coincidence probability is
\begin{equation}
\Pc=\frac34-\frac12\tr(\omega\tau_3).
\label{eq:P3}
\end{equation}
For a pair of identical pure photons, $\omega$ is their common state, and Eq.~\eqref{eq:P3} reduces to the result of Ref.~\cite{PradanaChew2019}, with $\tr(\omega\tau_3)$ in the role of the temporal overlap of the wave packets. As in the two-photon case, interference can only reduce the coincidences, in proportion to the overlap between the photon in one input and the photons in the other.

The combination of the four settings that defines the witness simplifies the analysis further. In each kind of event, the coincidence probability depends linearly on the state of a single photon that carries the setting. For two-photon events and for an extra photon in input $A$, this is the main photon $B$, in the state $\rho_j=\st j\rho\,\st j$. For an extra photon in input $B$, applying $\st j$ to the internal states of all three photons, which does not change the coincidence probability, since neither the BS nor threshold detectors act on the internal DOFs, transfers the setting to the main photon $A$. Since $\rho_1+\rho_2+\rho_3-\rho_0=2\Delta$, with $\Delta\equiv\Ip\otimes\rhoo-\rho$ [Eq.~\eqref{eq:twirl}], and since Eq.~\eqref{eq:WV} can be written as $\W_V=(C_1+C_2+C_3-C_0)/(2C_\infty)-1$, we obtain
\begin{equation}
\W_V=\frac{P_\Delta}{P_\infty}-1,
\label{eq:WVPD}
\end{equation}
where $P_\Delta$ is the coincidence probability that would be obtained, without any setting, if the photon that carries the setting were replaced by a photon in the state $\Delta$, and $P_\infty=\frac12(1-2\varepsilon)+\frac32\varepsilon=\frac12(1+\varepsilon)$ is that of the reference measurement, in which this photon is distinguishable from the others. If $\rho$ is separable, $\Delta$ is positive semidefinite by the reduction criterion, and, since $\tr\Delta=1$, it is a legitimate state. In two-photon events, $P_\Delta=\frac12[1-\tr(\rho\Delta)]\le\frac12$, and, in three-photon events, Eq.~\eqref{eq:P3} gives $P_\Delta=\frac34-\frac12\tr(\omega\Delta)\le\frac34$, where $\omega$ is the one-photon state of the two photons that share an input. Hence $P_\Delta\le P_\infty$ and $\W_V\le0$ for every separable $\rho$, whatever the states of the extra photons: a genuine photon cannot produce more coincidences by interfering with the other photons than by being distinguishable from them. In the absence of extra photons, Eq.~\eqref{eq:WVPD} gives back $\W_V=\W=-\tr(\rho\Delta)$.

With Eq.~\eqref{eq:P3}, the two models considered in the main text follow in one line each. If the extra photon is distinguishable from the main ones, i.e., occupies a mode of the other DOFs orthogonal to theirs, then $\omega=\frac12(\rho+\nu)$ and $\tr(\nu\Delta)=0$, so that the three-photon coincidence probability is $\frac34-\frac14\tr(\rho\Delta)$, and $\W_V=\W(1-\varepsilon)/(1+\varepsilon)$. If the extra photon is an independent copy of the main one, $\nu_A=\nu_B=\rho$, then $\omega=(\rho+\rho^2)/(1+p)$, with $p=\tr\rho^2$, and
\begin{equation}
\W_V=\frac{\W\,[1-2\varepsilon p/(1+p)]+2\varepsilon\W_T/(1+p)}{1+\varepsilon},
\end{equation}
where $\W_T=\tr[\rho^2(\rho-\Ip\otimes\rhoo)]=-\tr(\rho\Delta\rho)$ is a third-order moment that is nonpositive for separable states; for pure states, $p=1$ and $\W_T=\W$, so that $\W_V=\W/(1+\varepsilon)$. These expressions agree to machine precision with a direct second-quantization calculation of the three-photon interference, and numerical optimizations over separable states and arbitrary (mixed) extra-photon states, with $\varepsilon$ up to 0.2, gave no positive value of $\W_V$. Note that the relation $V\simeq \mathcal{F}-g^{(2)}(0)(1+\mathcal{F})$ between raw visibility and overlap found in the literature~\cite{Ollivier2021} refers to a different normalization of the coincidences and does not apply to Eq.~\eqref{eq:WV}.

%=====================================================================
\section{Qudits}\label{app:qudit}
%=====================================================================

With $X\ket k=\ket{k+1}$ (modulo $d$), $Z\ket k=\omega^k\ket k$, $\omega=e^{2\pi i/d}$, and $D_{ab}=X^aZ^b$, the Weyl-Heisenberg operators form a unitary 1-design~\cite{Gross2007}: for any operator $X$ acting on $\Hp\otimes\Ho$, where $\Hp$ is now $d$-dimensional,
\begin{equation}
\frac1{d^2}\sum_{a,b=0}^{d-1}(D_{ab}\otimes\Io)X(D_{ab}\otimes\Io)^\dagger=\frac{\Ip}{d}\otimes\tr_{\mathrm p}X ,
\end{equation}
which generalizes Eq.~\eqref{eq:twirl}. Applied to $X=\rho$, it gives $\sum_{a,b}\mathcal{F}(D_{ab})=d\tr\rhoo^2$, while $\mathcal{F}(I)=\tr\rho^2$; with $\mathcal{F}=1-2P$, this yields Eq.~\eqref{eq:Wd}, and the reduction criterion~\cite{HorodeckiReduction1999,Cerf1999} again gives $\W_d\le0$ for separable states. As in Appendix~\ref{app:proof}, the twirl implies $\rho\le d\,\Ip\otimes\rhoo$, so $\tr\rho^2\le d\tr\rhoo^2$, and, together with $\tr\rho^2\le1$, $\W_d\le\min[(d-1)\tr\rhoo^2,\,1-\tr\rhoo^2]\le1-1/d$; writing $\rhoo=\sum_k\rho_{kk}$ in terms of the diagonal blocks $\rho_{kk}=\bra k\rho\ket k$, the Cauchy-Schwarz inequality gives $\tr\rhoo^2\le d\sum_k\norm{\rho_{kk}}^2\le d\tr\rho^2$, so that $\W_d\ge-(1-1/d)\tr\rhoo^2\ge-(1-1/d)$. Any other unitary 1-design, which must have at least $d^2$ elements~\cite{Gross2007}, can replace the $D_{ab}$.

%=====================================================================
\section{Details of the simulations}\label{app:circuit}
%=====================================================================

\begin{figure*}[t]
\centering
\begin{tabular}{c@{\hspace{1.0cm}}c}
(a) & (b)\\[2pt]
\begin{quantikz}[row sep={0.62cm,between origins}, column sep=0.3cm, font=\small]
\lstick{$p_A$} & & \gate[3,style={fill=yellow!35}]{\mathcal P(q,D_0)} & & \gate{H}\gategroup[wires=8,steps=5,style={draw=none,fill=blue!8,rounded corners,inner xsep=2pt},background]{} & \swap{3} & & & & \\
\lstick{$t_A$} & & & & & & \swap{3} & & & \\
\lstick{$E_A$} & \qwbundle{3} & & & & & & & & \\
\lstick{$p_B$} & & \gate[3,style={fill=yellow!35}]{\mathcal P(q,D_0)} & \gate[style={fill=red!25}]{\sigma_j} & & \targX{} & & & & \\
\lstick{$t_B$} & & & & & & \targX{} & & & \\
\lstick{$E_B$} & \qwbundle{3} & & & & & & & & \\
\lstick{$\ket{0}_{\mathrm{aux}_1}$} & & & & \gate{H} & \ctrl{-3} & & \gate{H} & \meter{} & \setwiretype{c} \\
\lstick{$\ket{0}_{\mathrm{aux}_2}$} & & & & \gate{H} & & \ctrl{-3} & \gate{H} & \meter{} & \setwiretype{c}
\end{quantikz}
&
\begin{quantikz}[row sep={0.62cm,between origins}, column sep=0.45cm, font=\small]
\lstick{$p$} & & \gategroup[wires=3,steps=2,style={draw=none,fill=cyan!15,rounded corners,inner xsep=3pt},background]{} & \targ{} & \targ{}\gategroup[wires=5,steps=2,style={draw=none,fill=orange!20,rounded corners,inner xsep=3pt},background]{} & & \\
\lstick{$t$} & & \gate{R_Y(\vartheta)} & \ctrl{-1} & & \targ{} & \\
\lstick{$e_0$} & \gate[style={fill=yellow!35}]{R_Y(\phi)} & \octrl{-1} & \octrl{-1} & \ctrl{-2} & \ctrl{-1} & \\
\lstick{$e_1$} & \gate[style={fill=yellow!35}]{H} & & & \ctrl{-1} & & \\
\lstick{$e_2$} & \gate[style={fill=yellow!35}]{H} & & & & \ctrl{-2} &
\end{quantikz}
\\[4pt]
\multicolumn{2}{c}{%
\begin{tikzpicture}[every node/.style={font=\footnotesize}]
\node[anchor=west] (l1) at (0,0) {\tikz\fill[yellow!35] (0,0) rectangle (0.3,0.22); State preparation};
\node[anchor=west] (l2) at ([xshift=0.45cm]l1.east) {\tikz\fill[cyan!15] (0,0) rectangle (0.3,0.22); Internal entanglement};
\node[anchor=west] (l3) at ([xshift=0.45cm]l2.east) {\tikz\fill[orange!20] (0,0) rectangle (0.3,0.22); External entanglement};
\node[anchor=west] (l4) at ([xshift=0.45cm]l3.east) {\tikz\fill[red!25] (0,0) rectangle (0.3,0.22); Pauli operation};
\node[anchor=west] (l5) at ([xshift=0.45cm]l4.east) {\tikz\fill[blue!8] (0,0) rectangle (0.3,0.22); Swap tests};
\end{tikzpicture}}
\end{tabular}
\caption{(a) Quantum circuit employed in the simulations. Each photon $X\in\{A,B\}$ is represented by a polarization qubit $p_X$, a time-bin qubit $t_X$, and three environment qubits $E_X$, which are prepared by the block $\mathcal P(q,D_0)$ and never measured. After the Pauli operator $\sigma_j$ is applied to $p_B$, two controlled-swap tests compare the polarization and time-bin qubits of the two photons; the probability of obtaining ``01'' or ``10'' in the ancillas corresponds to the HOM coincidence probability. (b) Preparation block. The environment qubit $e_0$ is rotated so that it is found in $\ket0$ with probability $q$. If $e_0=0$ (open circles), the photon qubits are prepared in $\ket\Psi=\sqrt a\ket{00}+\sqrt b\ket{11}$, with $\cos^2(\vartheta/2)=a$ (internal entanglement). If $e_0=1$ (filled circles), $p$ and $t$ become maximally entangled with $e_1$ and $e_2$ (external entanglement). Tracing out the environment yields the state of Eq.~\eqref{eq:model}.}
\label{fig:circuit}
\end{figure*}
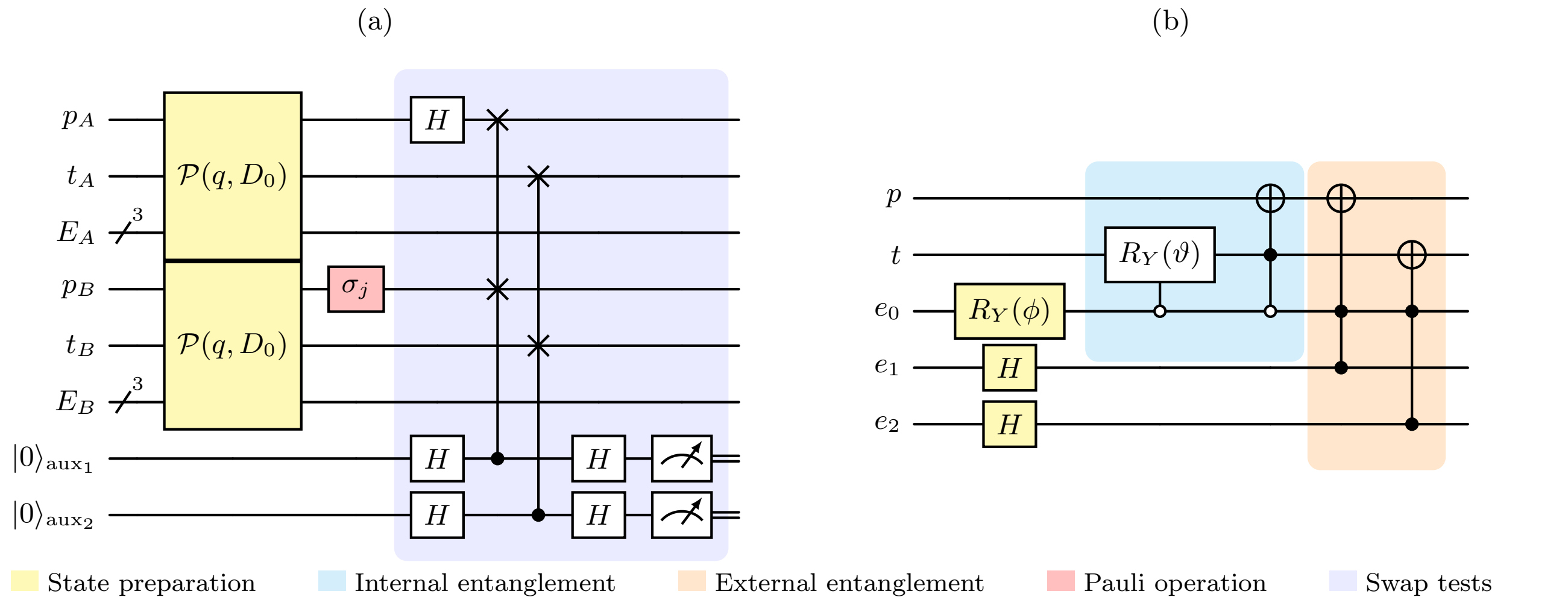

The quantum circuit is shown in Fig.~\ref{fig:circuit}(a). Each photon is represented by a polarization qubit ($p$) and a time-bin qubit ($t$), together with three environment qubits ($E$) that are never measured. The first environment qubit, $e_0$, acts as a ``coin'': with probability $q$ it is found in $\ket0$, and the photon qubits are prepared in the internally entangled state $\ket\Psi$, as in Fig.~5 of Ref.~\cite{Temporao2024}; with probability $1-q$ it is found in $\ket1$, and the polarization and time-bin qubits are each maximally entangled with one of the two other environment qubits, $e_1$ and $e_2$, as in Fig.~7 of Ref.~\cite{Temporao2024}. The two photons have independent environments. After the preparation, the Pauli operator $\sigma_j$ is applied to the polarization qubit of photon $B$, and two ancillas perform controlled-swap tests between the polarization qubits and between the time-bin qubits of the two photons.

In the preparation block of Fig.~\ref{fig:circuit}(b), the rotation $R_Y(\phi)$, with $\cos(\phi/2)=\sqrt q$, sets the weights of the two branches marked by $e_0$. For $e_0=0$, a controlled rotation $R_Y(\vartheta)$ on $t$, with $\cos^2(\vartheta/2)=a$, followed by a Toffoli gate that flips $p$, prepares $\ket\Psi=\sqrt a\ket{00}+\sqrt b\ket{11}$; for $e_0=1$, Toffoli gates entangle $p$ and $t$ maximally with $e_1$ and $e_2$, which are prepared in $\ket+$, leaving the photon qubits in $I_4/4$. Tracing out the environment yields Eq.~\eqref{eq:model}. The Pauli setting is applied as a $Z$, $X$, or $Y$ gate on $p_B$, corresponding to $\sigma_1$, $\sigma_2$, and $\sigma_3$, respectively. Each ancilla controls a swap between the corresponding qubits of the two photons; since the swap of a whole photon is the product of the swaps of its parts, the HOM coincidence probability equals the probability of odd parity of the two ancillas~\cite{Ekert2002,Foulds2021}. Simulations were performed with the \textsc{qiskit} Aer simulator~\cite{Qiskit,QiskitAer}, with $10^4$ shots per circuit and fixed random seeds. As an independent benchmark, we computed the exact output probabilities of all 336 circuits (4 values of $D_0$, 21 values of $q$, and 4 settings) from their state vectors, without sampling, and compared the probability of odd ancilla parity with a direct evaluation of Eq.~\eqref{eq:swap}, $\tr[(\rho\otimes\sigma_j\rho\sigma_j)\Pi_-]$, using the density operator of Eq.~\eqref{eq:model} and the swap operator, without any reference to the circuits. The largest difference was $1.0\times10^{-15}$. The normalized deviations of the sampled data of Fig.~\ref{fig:sim} with respect to Eq.~\eqref{eq:Wmodel} have mean $0.09$, root mean square $0.96$, and maximum absolute value $3.2$ over the 84 points. A script that recomputes these and all other numbers quoted in this paper, and checks them against the values given in the text, is included in the code archive (see Data availability).

Figure~\ref{fig:sim} shows the witness obtained for 21 values of $q$ and four values of $D_0$, with $10^4$ shots per Pauli setting. It changes sign at the threshold $q_{\W}$ and remains negative for every $q$ when $D_0=1$, in which case the photon is never internally entangled.

Concerning the random states used for witness sensitivity to mixed-state entanglement, density operators were generated as $\rho=GG^\dagger/\tr(GG^\dagger)$, where $G$ is a $4\times k$ matrix with independent complex Gaussian entries and $k$ is the rank. The negativity is $\mathcal N=\frac12(\norm{\rho^{T_{\mathrm p}}}_1-1)$, where $T_{\mathrm p}$ denotes the partial transpose with respect to polarization~\cite{Peres1996}.

%=====================================================================
\section{Quantum-dot model}\label{app:qdmodel}
%=====================================================================

We describe the photon by
$\rho=\sum_{a,b\in\{H,V\}}\ket a\!\bra b\otimes K_{ab}$,
where, in the time domain ($t,t'\ge0$),
\begin{multline}
K_{ab}(t,t')=c_ac_b\,\eta_{ab}\,\Gamma\,
e^{-\Gamma(t+t')/2}\\
\times e^{-i(\omega_at-\omega_bt')}
e^{-\gamma^*|t-t'|},
\end{multline}
with $c_H=\cos\theta$, $c_V=\sin\theta$,
$\omega_{H,V}=\omega_0\pm\Omega/2$, where $\Omega$
is the FSS in angular-frequency units
($S=\hbar\Omega$ in energy),
$\eta_{HH}=\eta_{VV}=1$, and
$\eta_{HV}=\eta_{VH}=\eta$.

The last factor describes Markovian pure dephasing
at rate $\gamma^*$, common to both transitions,
which therefore does not affect their equal-time
coherence. The parameter $\eta$ accounts for
the loss of coherence between the two bright states.
In particular, $\eta=0$ describes the unconditional
exciton-photon state in an ideal biexciton--exciton
cascade, after tracing out the biexciton photon.

For identical, independent photons,
$\W=\tr\rho^2-\tr\rhoo^2$, with
$\tr\rho^2=\sum_{a,b}\tr(K_{ab}K_{ba})$ and
$\rhoo=K_{HH}+K_{VV}$. The only integrals needed are
\begin{align}
\tr(K_{ab}K_{ba})
  &=Mc_a^2c_b^2\eta_{ab}^2,\\
\tr(K_{HH}K_{VV})
  &=\frac{Mc_H^2c_V^2}{1+\tilde\Omega^2},
\end{align}
where
\begin{equation}
M=\frac{\Gamma}{\Gamma+2\gamma^*},
\qquad
\tilde\Omega=\frac{\Omega}{\Gamma+2\gamma^*}.
\end{equation}
Here $\Gamma+2\gamma^*$ is the homogeneous
spectral linewidth (FWHM, in angular-frequency
units) of each polarization resolved transition.
These integrals yield
\begin{gather}
\tr\rho^2=
M\bigl[1-\tfrac12\sin^22\theta\,(1-\eta^2)\bigr],
\\
\tr\rhoo^2=
M\Bigl[1-\tfrac12\sin^22\theta\,
\frac{\tilde\Omega^2}{1+\tilde\Omega^2}\Bigr],
\end{gather}
and hence Eq.~\eqref{eq:Wqd}.

The time-integrated degree of polarization follows
from
$\tr K_{HV}=c_Hc_V\eta\Gamma/(\Gamma+i\Omega)$,
giving
\begin{equation}
\mathrm{DOP}=
\left[
\cos^22\theta+
\frac{\eta^2\sin^22\theta}{1+\Omega^2/\Gamma^2}
\right]^{1/2}.
\end{equation}
The closed-form expressions were checked against
a brute-force calculation in which $\rho$ is
discretized on a time grid, with agreement to
better than $10^{-4}$.

For $\theta=45^\circ$, three limiting cases are
instructive. When $\gamma^*=0$ and $\eta=1$,
the photon is pure and
\begin{equation}
\W=\frac{(\Omega/\Gamma)^2}
{2[1+(\Omega/\Gamma)^2]}.
\end{equation}
At $\Omega=0$, the state is a product state
with $\W=-M(1-\eta^2)/2$. Finally, for
$\Omega\gg\Gamma+2\gamma^*$, the witness
approaches $\W=M\eta^2/2$.

\emph{Complementary criterion.}
The global purity and the time-integrated DOP
also determine the complementary witness
$\W_{\mathrm p}$ of Eq.~\eqref{eq:Wp}.
For $\theta=45^\circ$, substitution of the
expressions above gives
\begin{equation}
\W_{\mathrm p}
=\frac M2(1+\eta^2)
-\frac12\left[
1+\frac{\eta^2}{1+\Omega^2/\Gamma^2}
\right].
\end{equation}
Unlike $\W$, this criterion becomes increasingly
sensitive to the reduction of global purity by
pure dephasing. In particular,
$\W_{\mathrm p}>0$ requires
$M(1+\eta^2)>1$, and is therefore impossible
for $M\le1/2$. By contrast, for $\eta=1$,
$\W>0$ for every $M>0$ and $\Omega>0$.

More importantly, the DOP and global purity do not
uniquely determine internal entanglement.
For example, the parameter sets
\begin{equation}
(\eta,\Omega/\Gamma,M)
=(1,\sqrt3,0.6)
\quad\text{and}\quad
(0.5,0,0.96)
\end{equation}
both give $\mathrm{DOP}=0.5$ and
$\tr\rho^2=0.6$, and hence the same
$\W_{\mathrm p}=-0.025$.
Nevertheless, their HOM witnesses are
$\W\simeq0.16$ and $\W=-0.36$, respectively.
The first state is internally entangled,
whereas the second is a product state.
Thus, measuring $\W$ provides information
that cannot be recovered from the DOP and
global purity alone.

\emph{Source characterization and spectral diffusion.}
Within the model, the sign of $\W$ is determined
by $\eta^2(1+\tilde\Omega^2)-1$.
The FSS, radiative lifetime, and
polarization-resolved HOM visibility determine
$\tilde\Omega$, while time-resolved polarimetry
can be used to estimate $\eta$.
These conventional characterizations therefore
allow the expected witness value to be predicted,
but the validity of a positive experimental
certificate does not depend on the model.

Slow spectral diffusion produces fluctuations
of the central frequency $\omega_0$.
When the two interfering photons experience
the same frequency shift, this corresponds to
the same unitary transformation on their
nonpolarization degrees of freedom and leaves
their HOM overlaps, and hence $\W$, unchanged.
This is relevant when the separation between
successive emissions is much shorter than
the correlation time of spectral diffusion.
In that case, however, the measured witness
characterizes the state conditioned on the
instantaneous central frequency, not necessarily
the ensemble-averaged single-photon state.
Independent frequency fluctuations between
the two photons require the more general
nonidentical-photon treatment discussed in
Sec.~\ref{sec:nonid}.

\emph{Time- and frequency-resolved polarimetry.}
Polarimetry with a time resolution much better
than $1/\Omega$, as used in time-resolved
studies of QD emission~\cite{Stevenson2008,Njala2026},
gives the polarization state conditioned on
the emission time $t$, obtained from the
diagonal elements $K_{ab}(t,t)$.
Its DOP,
\begin{equation}
\mathrm{DOP}(t)=
[\cos^22\theta+\eta^2\sin^22\theta]^{1/2},
\end{equation}
does not depend on $t$. Together with the
polarization populations, it determines $\eta$
within the model. Frequency-resolved
polarimetry provides analogous information.

Such measurements, however, cannot certify
internal entanglement on their own.
Any measurement resolving only the emission
time is described by operators diagonal in
the time basis, and therefore has identical
statistics for $\rho$ and for the state obtained
by discarding its off-diagonal temporal
coherences. In a discretized description,
the latter state has the form
\begin{equation}
\rho_{\mathrm{sep}}=
\sum_t \rho_{\mathrm p}^{(t)}
\otimes\ket t\!\bra t,
\end{equation}
where $\rho_{\mathrm p}^{(t)}$ is the
unnormalized polarization state conditioned
on $t$. This is a statistical mixture of
product states and is therefore separable.
The same argument applies to
frequency-resolved measurements in the
frequency basis.

\bibliography{refs}

\end{document}